\documentclass[showpacs,amsmath,amssymb,10pt,aps]{revtex4}

\usepackage{graphicx,color}
\usepackage{amsmath}
\usepackage{amssymb}
\usepackage{dsfont}
\usepackage{bbm}
\usepackage{color}
\usepackage{amsfonts}
\usepackage{bm}
\usepackage[most]{tcolorbox}
\usepackage{graphicx}
\usepackage{ulem}

\newtheorem{Proposition}{Proposition}
\newtheorem{theorem}{Theorem}

\newtheorem{Corollary}{Corollary}
\def\<{\langle}
\def\>{\rangle}
\usepackage[T1]{fontenc}
\usepackage[cp1250]{inputenc}

\begin{document}

\title{Eternally non-Markovian dynamics of a qubit interacting with a single-photon wavepacket}

\author{Anita D\c{a}browska}

\affiliation{Institute of Theoretical Physics and Astrophysics, University of Gda\'nsk, ul. Wita Stwosza 57, 80-308 Gda\'nsk, Poland}

\author{Dariusz Chru\'{s}ci\'{n}ski}

\affiliation{Institute of Physics, Faculty of Physics, Astronomy and Informatics,
Nicolaus Copernicus University, Grudziadzka 5/7, 87-100 Toru\'{n},
Poland}

\author{Sagnik Chakraborty}

\affiliation{Institute of Physics, Faculty of Physics, Astronomy and Informatics,
Nicolaus Copernicus University, Grudziadzka 5/7, 87-100 Toru\'{n},
Poland}

\author{Gniewomir Sarbicki}

\affiliation{Institute of Physics, Faculty of Physics, Astronomy and Informatics,
Nicolaus Copernicus University, Grudziadzka 5/7, 87-100 Toru\'{n},
Poland}



\begin{abstract}
An evolution of a two-level system (qubit) interacting with a single-photon wave packet is analyzed. It is shown that a hierarchy of master equations gives rise to phase covariant qubit evolution. The temporal correlations in the input field induce nontrivial memory effects for the evolution of a qubit. It is shown that in the resonant case whenever time-local generator is regular (does not display singularities) the qubit evolution never displays information backflow. However, in general the generator might be highly singular leading to intricate non-Markovian effects. A detailed analysis of the exponential profile is provided which allows to illustrate all characteristic feature of the qubit evolution.
\end{abstract}

\maketitle

\section{Introduction}

 Any real quantum system is never perfectly isolated and hence has to be treated as an open system \cite{BP02}. Therefore, its evolution is no longer unitary and gives rise to well known processes  of dissipation, decay, and decoherence induced by the nontrivial system--environment interaction \cite{Alicki}. When the interaction between system and environment is sufficiently weak and the characteristic time scales are well separated one uses well known Markovian approximation governed by the celebrated Markovian master equation $\dot{\rho} = \mathcal{L}\rho$, with $\mathcal{L}$ being the Gorini-Kossakowski-Lindblad-Sudarshan (GKLS) generator \cite{GKS,L}

\begin{equation}\label{GKLS}
  \mathcal{L}(\rho) = -i[H_{S},\rho] + \sum_k \gamma_k \Big( L_k \rho L_k^\dagger - \frac 12 \{ L_k^\dagger L_k,\rho\} \Big) ,
\end{equation}
where $H_{S}$ stands for the effective system's Hamiltonian (including Lamb shift correction),  $L_k$ are jump (Lindblad) operators, and $\gamma_k \geq 0$ (in what follows we keep $\hbar=1$). {In non-Markovian regime due to correlations between the system and environment the reduced evolution of the system is no longer governed by (\ref{GKLS}). One observes characteristic  memory effects such as information bacflow or recoherence. Laboratory techniques of quantum engineering enable nowadays to test experimentally theoretical descriptions and concepts of  non-Markovian quantum dynamics. Scope of applications of these theoretical tools is constantly growing and the field of non-Markovian quantum processes attracts a lot of interest (cf. recent reviews \cite{NM1,NM2,NM3,NM4} together with \cite{Piilo1,Piilo2,Modi2,Modi3} and the recent tutorial \cite{Modi4}).

In this paper we study non-Markovian  dynamics of the two-level system (qubit) interacting with a single-photon wave packet and the vacuum field. So we consider a scattering of a single-photon field on the two-level system. Propagating wave packets of light of definite numbers of photons play a key role in photonics. The continuous-mode single-photon state \cite{L00,M08} is an example of non-classical states of light having applications in quantum computation \cite{KLM01,Ralph03}, metrology \cite{D08,Munro02}, communication \cite{Scarani09}, and simulation \cite{Aaronon11}. Constantly developing techniques of producing, storing and detecting of single-photon states of light can serve in a wide range of practical applications  \cite{CWSS13,PSZ13,RR15,Leong2016,Zoller2017}.

We describe the evolution of qubit which is driven by the single-photon field and at the same time it undergoes damping process. Note that the single-photon field not only drives the system but it is also the source of stimulated emission. The key ingredient of our model are temporal correlations present in the input state of the field which is characterised by the time dependent profile $\xi(t)$ (for $t \geq 0$). The temporal correlations are responsible for all non-Markovian memory effects of the qubit dynamics. The reduced evolution of the qubit, obtained within the input-output formalism \cite{GarCol85}, is represented then by a hierarchy of coupled equations \cite{GEPZ98,WMSS11,BCBC12,Gough12a, Baragiola17,DSC17,DSC19,D20}. We provide the analytical solution to this hierarchy of equations for any initial state of the system and an arbitrary profile $\xi(t)$ and use it to show that the set of these equations is equivalent to a single time-local master equation. The price one pays for this reduction is highly nontrivial structure of time-dependent rates in the time-local master equation. The corresponding time-local generator is phase covariant and its general properties, recently studied, providing us with the necessary tools to analyze non-Markovanity effects \cite{SKHD16,Sabrina-2016,Sabrina-NJP,Francesco,HSH19,Sergey-2020}. The time dependent rates governing damping (cooling), heating and decoherence processes are fully characterized by the wave packet profile. Such formulae were never presented before. An immediate consequence of our analysis is the observation that in general the dynamical map governing the qubit evolution is not invertible which implies the singularity of rates in the corresponding time-local master equation. Similar observation for phase covariant dynamics was recently reported in \cite{Francesco}. Non-invertible maps were considered in quantum optics scenario, for instance, in \cite{Cresser1,Cresser2,Hou,SKHD16} (cf. also \cite{PRL-2018,CC19} and recent paper \cite{Jyrki-2021}).

The analyzed system provides excellent platform to test indicators of non-Markovianity. It turns out that in the resonant case,  which is of particular importance in photonic experiments, the evolution is never CP-divisible but it also never displays information backflow. We provide a detailed analysis of the qubit evolution interacting with exponential profile. In this case we are able to derive both analytical formulae for time-local rates and characterize the regime when the dynamics does not allow for information backflow.}

	The paper is organized  as follows. Section \ref{S-II} introduces  basic properties  of dynamical maps and discusses recent concepts of Markovianity. In Section \ref{S-III}  we introduce the hierarchy equations describing the reduced evolution of the two-level system interacting with the single-photon field and the vacuum, and eventually provide the solution for the qubit evolution. Sections \ref{SEC-IV} and \ref{SEC-V} contain the general properties of qubit evolution and the corresponding time-local generator, respectively. In Section \ref{EXP}  the detailed analysis of the exponential profile is carried out: we discuss both the resonant and off resonant case.
The key observation is that for an exponential profile the qubit evolution is eternally non-Markovian \cite{Erika}, that is, the corresponding time-local generator always contains a strictly negative  transition rate for any time $t > 0$. Actually, we conjecture that this characteristic feature holds for an arbitrary photon profile.  Final  conclusions are presented in  section \ref{S-VIII}. Technical details are presented in the appendices.

\section{Preliminaries}   \label{S-II}

Evolution of a quantum system is represented by a dynamical map $\{\Lambda_t\}_{t \geq 0}$ which maps an initial state $\rho$ at $t=0$ into a state at a current time $t$, i.e. $\rho(t) = \Lambda_t(\rho)$. Usually, the map is realized as a reduced evolution of the system + environment

\begin{equation}  \label{OPEN}
\rho(t)=\mathrm{Tr}_{E}\left(U(t)\rho\otimes \rho_{E}U^{\dagger}(t)\right),
\end{equation}
where $U(t)$ is the unitary evolution operator corresponding to the
total Hamiltonian $H=H_S+H_E+H_{int}$, and ${\rm Tr}_E$ denotes the partial trace over the environmental degrees of freedom \cite{BP02}. As a result (\ref{OPEN}) gives rise to a map $\rho \to \rho(t)$ which is completely positive and trace preserving (CPTP). Recall that any such map allows for the  Kraus representation

\begin{equation}\label{Kraus}
\Lambda_t(\rho) = \sum_{\alpha} K_\alpha(t) \rho K_\alpha^\dagger(t) ,
\end{equation}
where the time-dependent Kraus operator $K_\alpha(t)$ satisfies the additional normalization condition $\,\sum_{\alpha}  K_\alpha^\dagger(t) K_\alpha(t)= \mathbbm{1}\,$ which guaranties that the map $\Lambda_t$ is trace-preserving, i.e. ${\rm Tr}\, \Lambda_t(\rho) = {\rm Tr}\, \rho$ for any $\rho$. Reduced dynamics defined via (\ref{OPEN}) provides a generalization of Markovian dynamical semigroups governed by GKLS generator (\ref{GKLS}). For differentiable quantum dynamical map the system's density operator $\rho(t)$ satisfies time-local master equation
\begin{equation}\label{ME}
\dot{\rho}(t) = \mathcal{L}_t \rho(t) ,
\end{equation}
with time-dependent generator $\mathcal{L}_t$. Note, that given $\{\Lambda_t\}_{t \geq 0}$ the time-local generator is formally defined by $\,\mathcal{L}_t = \dot{\Lambda}_t \Lambda_t^{-1}$ provided the map $\Lambda_t$ is invertible. If the map is not invertible at $t=\tau$, then $\mathcal{L}_t$ displays singularities at $\tau$. This is not only a mathematical curiosity. Non-invertible maps were already considered in quantum optical  systems  \cite{Francesco,Cresser1,Cresser2,Hou,SKHD16} (cf. also \cite{PRL-2018,CC19} and recent papers \cite{Piilo1,Piilo2}). The system we study in this paper gives rise to a perfectly regular map  $\{\Lambda_t\}_{t \geq 0}$ which, however, needs not be invertible and hence the corresponding master equation is governed by a singular generator. Interestingly, singularities of $\mathcal{L}_t$ provide important physical insight into the dynamical properties of the system.

Any time-local generator $\mathcal{L}_t$ has essentially  the same structure as (\ref{GKLS}), where now the effective Hamiltonian $H_{S}$, the noise operators $L_k$, and the transition rates $\gamma_k$ are time-dependent. Given $\mathcal{L}_t$ the formal solution for the dynamical map reads as follows

\begin{equation}\label{T-exp}
\Lambda_t = \mathcal{T} \exp\left( \int_0^t \mathcal{L}_\tau d\tau \right) ,
\end{equation}
where $\mathcal{T}$ stands for the chronological product. It should be stressed that in the time-dependent case transition rates $\gamma_k(t)$ are not necessarily non-negative. This makes  the characterization of admissible generators extremely hard \cite{Erika,Angel}. Actually, temporal negativity of $\gamma_k(t)$ is usually interpreted as manifestation of non-Markovianity. A quantum dynamical map $\{\Lambda_{t}\}_{t \geq 0}$ is called divisible if for any $t \geq s$ one has $\,\Lambda_t = V_{t,s} \Lambda_s$. Note, that any invertible map is necessarily divisible since $V_{t,s} = \Lambda_t \Lambda_s^{-1}$. However, for not invertible maps the issue of divisibility is much more subtle \cite{BOGNA,datta,Acin-cor,Johansson,PRL-2018,CC19,Jyrki-2021}. In a recent paper \cite{Ujan} the construction  of propagators for non-invertible maps in terms of generalized inverse was proposed. One calls the map $\{\Lambda_t\}_{t\geq 0}$ P-divisible if $V_{t,s}$ is positive and trace-preserving, and CP-divisible if $V_{t,s}$ is completely positive and trace-preserving. Following  \cite{RHP} we call the evolution represented by $\{\Lambda_t\}_{t\geq 0}$ Markovian if it is CP-divisible. An invertible map is CP-divisible if and only if all rates satisfy $\gamma_k(t) \geq 0$. This is a direct generalization of Markovian semigroup. For non-invertible maps the full characterization is still missing \cite{PRL-2018}. One of the most interesting implication of non-Markovianity is an information backflow defined as follows \cite{BLP}: if the evolution is Markovian, then for any pair of initial states $\rho_1$ and $\rho_2$
\begin{equation}\label{BLP}
\frac{d}{dt}\| \Lambda_t(\rho_1 - \rho_2) \|_1 \leq 0 ,
\end{equation}
where  $\| X \|_1= {\rm Tr}|X|$ denotes the trace norm of $X$. Note, that the quantity $\|\rho_1 - \rho_2 \|_1 $ describes the distinguishability of $\rho_1$ and $\rho_2$. Now, the violation of (\ref{BLP}) is interpreted as an information backflow (from the environment to the system). In what follows we call (\ref{BLP}) a BLP condition. An interesting example of the qubit evolution which satisfies BLP condition but is not CP-divisible was provided in \cite{Erika}: the corresponding time-local generator reads as follows

\begin{equation}\label{}
\mathcal{L}_t(\rho) = \frac 12 \sum_{k=1}^3 \gamma_k(t)( \sigma_k \rho \sigma_k - \rho) ,
\end{equation}
with $\gamma_1=\gamma_2=1$ and $\gamma_3(t) = - \tanh t < 0$. Interestingly, one of the rate is always negative (for $t>0$) but still the corresponding dynamical map is completely positive. The authors of \cite{Erika} call such evolution {\it eternally non-Markovian}.  This analysis was further generalized in \cite{Nina} and  \cite{Kasia} for qudit systems. In this paper we face another example of qubit evolution for which for any $t>0$ some rate $\gamma_k(t)$ is negative. Nevertheless, the evolution is represented by completely positive dynamical map.
This might be, therefore, considered as an another example of eternally non-Markovian evolution. This example, however, corresponds to well motivated physical system representing a qubit interacting with an incoming external photon.

It should be stressed that dealing with evolution of open quantum systems  the very notion of {\it Markovian evolution} is often used in different contexts and actually several concept of (non)Markovian evolution are available in the current literature (cf. recent reviews \cite{NM1,NM2,NM3,NM4} and \cite{Piilo1,Piilo2}). Authors of \cite{NM4} shown an intricate hierarchy of various notions of quantum Markovianity and clearly indicates that this concept is highly context-dependent. In the present paper we are characterizing non-Markovian process using only the properties of the corresponding dynamical map. An interesting  approach to quantum non-Markovianity beyond a dynamical map based on so called quantum process tensor was recently analyzed in \cite{Modi2,Modi3} (cf. also recent tutorial \cite{Modi4}). This approach was further applied in  \cite{Andrea1,Andrea2,Filippov-Modi} in the study of open quantum systems dynamics.

A simple indicator of non-Markovianity was proposed in  \cite{LPP13}   (see also further discussion in \cite{CMM17}): for any Markovian evolution

\begin{equation}\label{GEO}
\frac{d}{dt} |{\rm det}\Lambda_t| \, \leq\,  0 .
\end{equation}
The volume of the accessible states at time $t$ is given by $\mathrm{Vol}(t)= |{\rm det}\Lambda_t| \rm{Vol}(0)$ and hence (\ref{GEO}) implies monotonic decrees of $\rm{Vol}(t)$. Any P-divisible (and hence also CP-divisible) satisfies (\ref{GEO}). In what follows we analyze a qubit interacting with  an incoming photon field  and show how the photon's profile $\xi(t)$ decides about the validity of BLP condition (\ref{BLP}), geometric condition (\ref{GEO}), and eventually (non)Markovianity of the evolution.


\section{Hierarchy of master equations}   \label{S-III}

\begin{figure}
	\includegraphics[width=5cm]{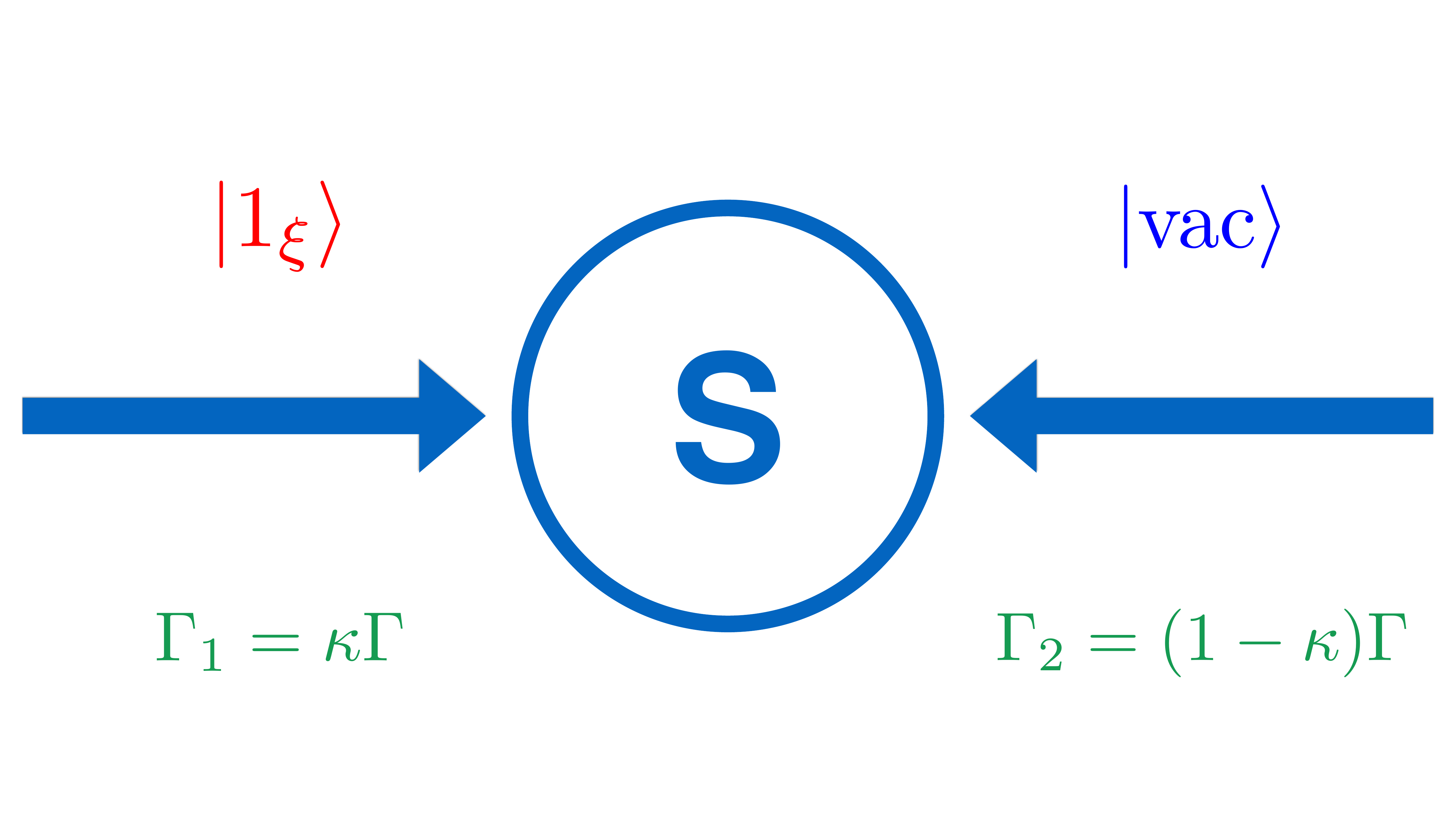}
	\caption{The system S interacts from the left with a single-photon field $|1_\xi\>$ and from the right with a vacuum field $|vac\>$. We introduced $\Gamma = \Gamma_1 + \Gamma_2$ and the parameter $\kappa \in [0,1]$ such that $\Gamma_1 = \kappa \Gamma$, and $\Gamma_2 = (1-\kappa) \Gamma$. } \label{Pic}
\end{figure}

We consider a two-level system interacting by two channels with a continuous-mode electromagnetic field \cite{L00} schematically depicted in  Fig. \ref{Pic}. The system interacts with the single-photon field incoming from the left and the vacuum field incoming from the right.
We assume that, in general, the coupling of the system with the single-photon field and the vacuum could be different. In the model some standard assumptions in quantum optics are made (rotating wave-approximation, a flat coupling constant, and an extension of the lower limit of integration over frequency to minus infinity). The bandwidth of the spectrum is assumed to be much smaller that the central frequency of the pulse $\Delta \omega \ll \omega_{c}$ and the central frequency $\omega_c$ is taken to be close the transition frequency of the two-level system $\omega_{0}$. A detailed discussion on the assumptions for the model one can find, for instance, in \cite{FTVRS17}. We assume that the composed system is initially in the state $\rho(0)\otimes |1_{\xi}\rangle\langle 1_{\xi}|\otimes |vac \rangle\langle vac|$, where
\begin{equation}
|1_{\xi}\rangle=\int_{-\infty}^{+\infty} d\omega \hat{\xi}(\omega)a^{\dagger}_{1}(\omega)|vac\rangle
\end{equation}
and $|vac\rangle$ is the continuous-mode vacuum state, $a^{\dagger}_{i}(\omega)$, $a_{i}(\omega)$ $(i=1,2)$ are annihilation and creation operators of $\omega$-mode satisfying the standard canonical commutation relations
\begin{equation}
[a_{i}(\omega),a^{\dagger}_{j}(\omega^{\prime})]=\delta_{ij}\delta(\omega-\omega^{\prime}).
\end{equation}
The profile of the photon in the time domain is defined as follows
\begin{equation}
\xi(t)=\int_{-\infty}^{+\infty} d\omega \hat{\xi}(\omega)e^{-i(\omega-\omega_{c})t}
\end{equation}
with the normalization condition
\begin{equation}
\int_{0}^{+\infty} dt|\xi(t)|^2 = \int_{-\infty}^{+\infty} d\omega |\hat{\xi}(\omega)|^2 = 1.
\end{equation}
Note that the profile, $\xi(t)$, defines a slow varying envelope of the time amplitude of the pulse. Since we are close to the resonance $(\omega_c \approx \omega_0$) it is convenient to pass to the interaction picture defined by the central frequency $\omega_c$, that is, a qubit state $\rho(t)$ `rotates' via $e^{-i \omega_c \sigma_z/2}$. The interaction Hamiltonian in the interaction-picture has the form
\begin{equation}
H_{int}(t)=i\sqrt{\Gamma_{1}}\left(\sigma_{-}a^{\dagger}_{1}(t)-\sigma_{+}a_{1}(t)\right)+i\sqrt{\Gamma_{2}}\left(\sigma_{-}a_{2}^{\dagger}(t)-\sigma_{+}a_{2}(t)\right),
\end{equation}
where $a_{i}(t)$ and $a^{\dagger}_{i}(t)$ are quantum white-noise operators \cite{GZ10} obeying the communication relation
\begin{equation}
[a_{i}(t),a_{j}^{\dagger}(t^{\prime})]=\delta_{ij}\delta(t-t^{\prime}) .
\end{equation}
Using the `rotating frame' one derives  the following hierarchy of coupled equations for the family of qubit operators ${\varrho}^{kl}$ with $k,l=0,1$
\begin{equation}\label{mast1}
\dot{\varrho}^{11}(t)=\mathcal{L}{\varrho}^{11}(t)
+\sqrt{\Gamma_{1}}\;\xi^{\ast}(t)[ \sigma_{-}, {\varrho}^{10}(t)]-
\sqrt{\Gamma_{1}}\;\xi(t)[\sigma_{+}, {\varrho}^{01}(t) ],
\end{equation}
\begin{equation}\label{mast2}
\dot{{\varrho}}^{10}(t)= \mathcal{L}{\varrho}^{10}(t)-
\sqrt{\Gamma_{1}}\;\xi(t)\left[\sigma_{+}, {\varrho}^{00}(t)\right],
\end{equation}
\begin{equation}\label{mast3}
\dot{{\varrho}}^{01}(t)= \mathcal{L}{\varrho}^{01}(t)+
\sqrt{\Gamma_{1}}\;\xi^{\ast}(t)\left[\sigma_{-},{\varrho}^{00}(t)\right],
\end{equation}
\begin{equation}\label{mast4}
\dot{{\varrho}}^{00}(t)= \mathcal{L}{\varrho}^{00}(t)
\end{equation}
with the super-operator
\begin{equation}
\mathcal{L}({\varrho})=-\frac{i\Delta_{0}}{2}[ {\varrho},\sigma_{z}]-
\frac{\Gamma}{2}\left\{\sigma_{+}\sigma_{-},{\varrho}\right\}
+\Gamma\sigma_{-}{\varrho}\sigma_{+},
\end{equation}
where $\Delta_{0}=\omega_{c}-\omega_{0}$, $\sigma_{z}=|e\rangle \langle e|-|g\rangle \langle g|$, $\Gamma=\Gamma_{1}+\Gamma_{2}$, and $\Gamma_{1},\Gamma_{2}\geq 0$. By $|e\rangle$ and $|g\rangle$ the excited and ground states of the two-level system are respectively denoted. One defines the density operator of the system  $\rho(t) := \varrho^{11}(t)$, and initially
\begin{equation}\label{}
\varrho^{11}(0) =  \varrho^{00}(0) = \rho(0) \ , \ \  \varrho^{01}(0) =  \varrho^{10}(0) = 0 .
\end{equation}
Note, that $\varrho^{10}(t)=\left(\varrho^{01}(t)\right)^{\dagger}$. The system is initially in an arbitrary state, $\rho(0)$. The set of equations defines the qubit dynamics during the scattering process. 
The derivations of the set of equations describing the evolution of a quantum system interacting with the single-photon field was given, for instance, in \cite{GEPZ98,BCBC12,WMSS11,Gough12a,Baragiola17}. The details of its determination in a collision model \cite{Ciccarello2021} one can find in \cite{DSC17}. Let us notice that the above set of equations can be written in the form
\begin{equation}\label{HHH}
  \frac{d}{dt} \begin{pmatrix}
                 \varrho^{11}(t) \\
                 \varrho^{10}(t) \\
                 \varrho^{01}(t) \\
                 \varrho^{00}(t)
               \end{pmatrix} = \begin{pmatrix}
                                 \mathcal{L} &  \mathcal{L}_{\xi} & \mathcal{L}_\xi^\ddag &0 \\
                                 0 & \mathcal{L} & 0 & \mathcal{L}_{\xi}^\ddag\\
                                 0 & 0 & \mathcal{L} & \mathcal{L}_{\xi} \\
                                 0 & 0 & 0 & \mathcal{L}
                               \end{pmatrix} \begin{pmatrix}
                 \varrho^{11}(t) \\
                 \varrho^{10}(t) \\
                 \varrho^{01}(t) \\
                 \varrho^{00}(t)
               \end{pmatrix} ,
\end{equation}
where  we introduced
\begin{equation}\label{}
  \mathcal{L}_\xi(X) = \sqrt{\Gamma_{1}}\; \xi^*(t) [\sigma_-,X] ,
\end{equation}
and its dual
\begin{equation}\label{}
  \mathcal{L}^\ddag_\xi(X) = - \sqrt{\Gamma_{1}}\;\xi(t) [\sigma_+,X] ,
\end{equation}
where ${\rm Tr}( \mathcal{L}^\ddag_\xi(X) Y) = {\rm Tr}(X \mathcal{L}_\xi(Y))$ for any pair $X,Y$.

Note, that operators  $\varrho^{01}(t)$ and $\varrho^{10}(t)$ are not Hermitian. However, introducing a pair of Hermitian operators

\begin{equation}\label{}
  \varrho^2(t) := \varrho^{01}(t) + \varrho^{10}(t) \ , \ \ \  \varrho^3(t) := -i(\varrho^{01}(t) - \varrho^{10}(t)) \ ,
\end{equation}
together with $ \varrho^1(t) := \varrho^{11}(t)$ and $ \varrho^4(t) := \varrho^{00}(t)$, the systems of equations (\ref{HHH}) may be rewritten as follows

\begin{eqnarray} \label{HHHH}
  \dot{\varrho}^1(t)  &=& \mathcal{L} \varrho^1(t) - \frac{i}{2} \sqrt{\Gamma_1}\, \Big( [\sigma^\xi_y,\varrho^2(t)]+ [\sigma^\xi_x,\varrho^3(t)] \Big) , \nonumber \\
  \dot{\varrho}^2(t)  &=& \mathcal{L} \varrho^2(t) - i \sqrt{\Gamma_1}\,  [\sigma^\xi_y,\varrho^4(t)] , \nonumber \\
   \dot{\varrho}^3(t)  &=& \mathcal{L} \varrho^3(t) - i \sqrt{\Gamma_1}\,  [\sigma^\xi_x,\varrho^4(t)] ,  \\
  \dot{\varrho}^4(t)  &=& \mathcal{L} \varrho^4(t) , \nonumber
\end{eqnarray}
where

\begin{equation}\label{}
  \sigma^\xi_x := \xi^* \sigma_- + \xi \sigma_+ = \left( \begin{array}{cc} 0 & \xi^* \\ \xi & 0 \end{array} \right) ,
\ \ \ \
  \sigma^\xi_y := i(\xi^* \sigma_- - \xi \sigma_+) = \left( \begin{array}{cc} 0 & i\xi^*  \\  -i \xi & 0 \end{array} \right) .
\end{equation}

The set of equations (\ref{HHHH}) has a structure of  hierarchical master equations of
motion (HEOM)  \cite{Florian,Tanimura}	
\begin{equation}\label{F}
  \dot{\varrho}^i(t) = \sum_{j=1}^n \mathcal{L}_{ij} \varrho^j(t) \ , \ \ i,j=1,\ldots,n ,
\end{equation}
for some $n \geq 1$. In Eq. (\ref{F}) only $\varrho^1$ is considered as a system's state and the remaining operators $\varrho^i$ are auxiliary objects. For the system (\ref{HHHH}) one has {$n=2^2$}. Interestingly,  in a more general case, when the field is prepared in $N$-photon state, the evolution of the system is governed by the set of $(N + 1)^2$ coupled equations \cite{BCBC12,DSC19}.

Note, that if $\xi(t)=0$, i.e. there is no photon, then (\ref{HHHH}) reduces to

\begin{equation}\label{}
  \dot{\varrho}^k(t)  = \mathcal{L} \varrho^k(t) \ , \ \ \ k=1,2,3,4 ,
\end{equation}
and the auxiliary operators $\{\varrho^2,\varrho^3,\varrho^4\}$ are totally decoupled from the evolution of the system itself. The photon profile $\xi(t)$ makes the systems (\ref{HHHH}) nontrivial and eventually influences the evolution of $\rho(t) := \varrho^1(t)$. As we shall see such evolution is no longer Markovian.

\section{General properties of the qubit dynamics}   \label{SEC-IV}

The hierarchy of equations (\ref{HHH}) may be solved  (cf. Appendix \ref{AppenA}) and eventually one finds the following dynamical map $\Lambda_t$ representing qubit evolution \cite{D20}

\begin{equation}\label{rho-t}
\rho(0) \ \to \    {\rho}(t) = \Lambda_t \rho(0) = \left( \begin{array}{cc}1-P_{e}(t) & C(t)  \rho_{ge}(0) \\
C^{\ast}(t) \rho_{eg}(0)  & P_{e}(t) \\
\end{array} \right),
\end{equation}
where the  population of the excited state $P_e (t) = \rho_{ee}(t)$ reads as follows
\begin{equation}
P_{e}(t)=A(t)+B(t)P_{e}(0),
\end{equation}
together with
\begin{equation}\label{A(t)}
A(t)=\kappa\Gamma e^{-\Gamma t} \left|\int_{0}^{t}ds \xi(s)e^{\left(-i\Delta_{0}+\frac{\Gamma}{2}\right) s}\right|^2,
\end{equation}
\begin{equation}\label{B(t)}
B(t)=e^{-\Gamma t}\left(1-	4\kappa\Gamma  \mathrm{Re}\int_{0}^{t}ds\xi^{\ast}(s)e^{\left(i\Delta_{0}+\frac{\Gamma}{2}\right)s}\int_{0}^{s}d\tau\xi(\tau)e^{\left(-i\Delta_{0}-\frac{\Gamma}{2}\right)\tau}\right),
\end{equation}
and
\begin{equation}  \label{C(t)}
C(t)=e^{\left(-i\Delta_{0}-\frac{\Gamma}{2}\right)t}\left(1-2\kappa\Gamma \int_{0}^{t}ds\xi^{\ast}(s)e^{\left(i\Delta_{0}-\frac{\Gamma}{2}\right) s}\int_{0}^{s}d\tau\xi(\tau)e^{\left(-i\Delta_{0}+\frac{\Gamma}{2}\right) \tau}\right).
\end{equation}
We introduce here the parameter $\kappa= \Gamma_{1}/\Gamma$, hence $\kappa\in [0,1]$. Thus, if $\kappa=0$, we deal with the system interacting only with the vacuum part, and if $\kappa=1$, we observe only an interaction with the single-photon field. The above formulae provide a complete description of the qubit evolution for an {\it arbitrary} photon profile $\xi(t)$.
Note, that $B(t)$ together with $\{C(t),C^*(t)\}$ define time dependent eigenvalues of the dynamical map $\Lambda_t$:

\begin{equation}\label{eigen}
  \Lambda_t(|e\>\<e| - |g\>\<g|) = B(t) (|e\>\<e| - |g\>\<g|)\ ,  \ \ \Lambda_t|g\>\<e|= C(t) |g\>\<e| \ , \ \ \Lambda_t|e\>\<g|= C^*(t) |e\>\<g| .
\end{equation}
 Clearly, the evolution highly depends upon the photon profile and the dependence of $\{A(t),B(t),C(t)\}$ upon $\xi(t)$ is quite nontrivial. However, the asymptotic state is universal: the atom eventually relaxes to the ground state $|g\>$ irrespective of $\xi(t)$. It follows from the asymptotic analysis

\begin{equation}
\lim_{t\to +\infty}A(t)=0,\;\;\lim_{t\to +\infty}B(t)=0,\;\;\lim_{t\to +\infty}C(t)=0.
\end{equation}
Hence, we obtain $\displaystyle{\lim_{t\to +\infty}}P_{e}(t)=0$ and $\displaystyle{\lim_{t\to +\infty}}\rho(t)=|g\rangle\langle g|$. It should be stressed, however, that $|g\>\<g|$ is not an invariant state of the evolution. This is essential difference between Markovian semigroup and non-Markovian evolution \cite{PRA-Saverio}. For a semigroup an asymptotic state always defines an invariant state. However, for the evolution (\ref{rho-t}) if one starts with $\rho(0)=|g\>\<g|$, then

\begin{equation}\label{eAg}
  \rho(t) = A(t)|e\>\<e| + [1-A(t)] |g\>\<g| ,
\end{equation}
that is, the evolution does not create coherence but it does create population controlled by $A(t)$ which asymptotically vanishes.
The problem of an optimal profile, i.e. a profile maximizing $A(t)= P_e(t)$ for some finite $t$,  was analyzed by several authors \cite{SAL10a,Rephali2010,WMSS11,RB17}. Our analysis allows to formulate the following

\begin{Proposition} \label{PRO-1} If $\rho(0) = |g \rangle\langle g|$, then the maximal population of the excited state $P_e(t)$ at time $t>0$ reads as follows
\begin{equation}\label{A-max}
P^{\rm max}_e(t) :=   \max_{\xi(t)} P_e(t) = \kappa\left(1 - e^{-\Gamma t}\right),
\end{equation}
and is realized only at the resonance (i.e. $\Delta_{0} =0$)  by the following pulse
\begin{equation}\label{xi-max}
  \xi(\tau) = \sqrt{ \frac{\Gamma}{e^{\Gamma t}-1} } \, e^{\frac{\Gamma}{2}\tau}
\end{equation}
for $\tau \in[0,t]$, and $\xi(\tau)=0$ for $\tau > t$.
\end{Proposition}
For the proof cf. Appendix \ref{AppenB}. Let us note that for any given profile, $\xi(t)$, one may discuss the problem of choosing its parameters which provides the maximal temporal excitation of the system \cite{WMSS11,RB17}.

 We conclude this section with the following statements. There are two essentially different scenario for the qubit evolution:

\begin{enumerate}
  \item The quantities $B(t) > 0$ and $|C(t)| > 0$ {for all $t\geq 0$.} In this case the dynamical map $\Lambda_t$ is invertible (eigenvalues $\{B(t),C(t),C^*(t)\}$ do not vanish), that is, knowing a qubit state $\rho(t)$ at time $t$ one may reconstruct the initial state $\rho(0)$.

  \item The dynamical map $\Lambda_t$ is no longer invertible, i.e. there exists $t < \infty$ such that either $B(t)=0$ or $C(t)=0$. Note, that whenever $B(t)=0$,  the density operator $\rho(t)$ reads as
\begin{equation}\label{}
{\rho}(t)=\left( \begin{array}{cc}1- A(t) & C(t)  \rho_{ge}(0) \\
C^{\ast}(t) \rho_{eg}(0)  & A(t) \\
\end{array} \right),
\end{equation}
and hence it does not depend upon initial population $P_e(0)$. Actually, diagonal elements display the same structure as in (\ref{eAg}). So if initially $\rho_{ge}(0)=(\rho_{eg}(0))^{\dagger}=0$, then all trajectories of different $P_{e}(0)$ cross at some point. Whenever $C(t)=0$, then
\begin{equation}   \label{}
{\rho}(t)=\left( \begin{array}{cc}1- P_e(t) & 0 \\
0  & P_e(t) \\
\end{array} \right),
\end{equation}
and for this moment the information about qubit coherence is lost.
\end{enumerate}

\section{Time-local generator and non-Markovianity conditions}   \label{SEC-V}

{To discuss non-Markovianity of qubit  evolution it is convenient to introduce the corresponding time-local master equation for the density matrix. The formula (\ref{rho-t}) for time evolution of the qubit density operator $\rho(t)$ defines a dynamical map $\rho(0) \to \Lambda_t \rho(0)$. By differentiating (\ref{rho-t}) with respect to time, one easily finds that $\Lambda_t$ satisfies time-local master equation $\dot{\Lambda}_t = \mathcal{L}_t \Lambda_t$, with the following time-local generator

\begin{equation}\label{L}
\mathcal{L}_t(\rho) = - i\frac{\omega(t)}{2} [\sigma_z,\rho] + \frac{\gamma_+(t)}{2} \left(\sigma_+ \rho \sigma_- - \frac 12 \{ \sigma_-\sigma_+,\rho\} \right) + \frac{\gamma_-(t)}{2} \left(\sigma_- \rho \sigma_+ - \frac 12 \{ \sigma_+\sigma_-,\rho\} \right) + \frac{ \gamma_z(t)}{2} (\sigma_z\rho \sigma_z -\rho) ,
\end{equation}
where $\sigma_{\pm}=\frac{1}{2}(\sigma_{x}\pm i\sigma_{y})$, and $\sigma_{x}$, $\sigma_{y}$, $\sigma_{z}$ are the Pauli operators. The real valued time-dependent rates $\gamma_{+}(t)$, $\gamma_{-}(t)$, and $\gamma_{z}(t)$, describing, respectively, pumping, damping and pure dephasing, are defined as follows
\begin{equation}\label{gg}
  \gamma_+(t) = 2 \frac{\dot{A}(t)B(t) - A(t)\dot{B}(t)}{B(t)} \ , \ \ \
  \gamma_-(t) = -2 \frac{\dot{B}(t)}{B(t)} - \gamma_+(t) ,
\end{equation}

\begin{equation}\label{gz}
\gamma_z(t) = \frac 12 \frac{\dot{B}(t)}{B(t)} - \frac{\frac{d}{dt}{|C(t) |}}{|C(t)|} = \frac 12 \frac{\dot{B}(t)}{B(t)} - {\rm Re}\, \frac{\dot{C}(t)}{C(t)}   ,
\end{equation}
and
\begin{equation}
  \omega(t) = \frac{\dot{C}(t)|C(t)|-C(t)\frac{d}{dt}|C(t)|}{iC(t)|C(t)|} =  {\rm Im}\, \frac{\dot{C}(t)}{C(t)}.
\end{equation}
It should be emphasized that time-local generator $\mathcal{L}_t$ becomes singular whenever $B(t)=0$ or $C(t)=0$. In this case $\gamma_\pm(t)$ and $\gamma(t)$ diverge. Nevertheless, the dynamical map $\Lambda_t = \mathcal{T} \exp( \int_0^t \mathcal{L}_u du)$ is always well defined for any $t \geq 0$. It shows that perfectly regular hierarchy of the equations (\ref{mast1})-(\ref{mast2}) may give rise to singular master equation for a system's density operator $\rho(t)$. One can easily check that if  $\xi(t)=0$ for any $t\geq 0$, then we get $   A(t) = 0,\, B(t) = e^{-\Gamma t}, \, C(t) = e^{-\left(i\Delta_{0} + \Gamma/2\right)t}$,
	and $  \gamma_+(t) = 0, \, \gamma_-(t) =  2\Gamma, \, \gamma_z(t) =0, \, \omega(t) =  \Delta_{0}$, that is, $\mathcal{L}_t = \mathcal{L}$.

Interestingly, as shown in \cite{SKHD16}, (\ref{L}) provides  the most general time-local generator which satisfies the following covariance property

\begin{equation}\label{phase}
  U_{\varphi} \mathcal{L}_t (\rho)  U^\dagger_{\varphi} = \mathcal{L}_t ( U_{\varphi}\rho U^\dagger_{\varphi}) ,
\end{equation}
for any diagonal unitary operator
\begin{equation}\label{}
  U_{\varphi} = e^{i\varphi } |g\rangle \langle g| + e^{-i\varphi} |e\rangle \langle e| ,
\end{equation}
which means that the map $\Lambda_{t}$ commutes with the free evolution of the system.

Conditions for Markovianity, as well as positivity and complete positivity of phase covariant {qubit} dynamical map have been extensively studied recently  \cite{Sabrina-2016,Sabrina-NJP,Sergey-2020}. Clearly, if $\gamma_{\pm}(t)\geq 0$ and $\gamma_{z}(t)\geq 0$ for all $t\geq 0$, then the evolution is CP-divisible.  However, $\gamma_{\pm}(t)$ and $\gamma_{z}(t)$ can be temporally negative without violating CP condition for $\Lambda_{t}$. {Of course, by construction the evolution defined by (\ref{mast1})-(\ref{mast4}) is always CP.} For the time-local generator (\ref{L}) one can provide the following characterization \cite{Sabrina-NJP,Sergey-2020}.

\begin{theorem}  \label{TH} The qubit evolution generated by (\ref{L})
	
	\begin{itemize}
		\item is CP-divisible iff $\gamma_\pm(t) \geq 0$ and $\gamma_z(t) \geq 0$,
		\item is P-divisible iff
		\begin{equation}\label{}
		\gamma_\pm(t) \geq 0 , \ \ \ \sqrt{\gamma_+(t)\gamma_-(t)} + 2 \gamma_z(t) \geq 0 ,
		\end{equation}
		\item satisfies BLP condition iff
		\begin{equation}\label{BLP1}
		\gamma_+(t) + \gamma_-(t)  \geq 0 , \ \ \ \gamma_+(t) + \gamma_-(t) + 4 \gamma_z(t)  \geq 0 ,
		\end{equation}	
		\item satisfies a  geometric criterion (\ref{GEO}) iff 	
		\begin{equation}\label{GEO1}
		\gamma_+(t) + \gamma_-(t) + 2 \gamma_z(t) \geq 0 ,
		\end{equation}
	\end{itemize}
	for all $t \geq 0$.
\end{theorem}
Note that the BLP as well as the geometric condition are fully controlled by $B(t)$ and $C(t)$, only. One can easily check that
\begin{eqnarray}\label{}
  \gamma_+(t) + \gamma_-(t) = - 2 \frac{\dot{B}(t)}{{B(t)}} \ , \ \ \  \gamma_+(t) + \gamma_-(t) + 4 \gamma_z(t) = -4  {\rm Re}\, \frac{\dot{C}(t)}{C(t)}.
\end{eqnarray}
\begin{eqnarray}\label{}
\gamma_+(t) + \gamma_-(t) + 2 \gamma_z(t) = - \frac{\dot{B}(t)}{{B(t)}} -2 {\rm Re}\, \frac{\dot{C}(t)}{C(t)}.
\end{eqnarray}
It clearly shows the difference between BLP condition and P-divisibility which is more demanding and depends also upon $A(t)$. Detailed properties of qubit evolution depend upon the profile $\xi(t)$. Assuming resonance we can provide the detailed analysis for a large class of profiles. For the profile $\xi(t)\in \mathbb{R}_{+}$ the evolution is never CP-divisible. However, for invertible dynamical map the weaker condition for no information backflow is always satisfied.

\begin{Proposition}\label{PRO-2} Let $\Delta_{0}=0$ and $\xi(t)\in \mathbb{R}_{+}$.  If $B(t)>0$  for all $t > 0$, then there is no information backflow (BLP condition holds). One has $C(t)>0$ for all $t > 0$ and the coherence, $C(t)|\rho_{ge}(0)|$, decreases monotonically in the course of time. Moreover,

\begin{equation}\label{}
  \gamma_+(t) \geq 0 \ , \ \ \gamma_-(t) \geq 2\Gamma \ , \ \ \gamma_z(t) \leq 0 ,
\end{equation}
and hence the evolution is eternally non-Markovian.
	
\end{Proposition}
For the proof see Appendix \ref{AppenC}. In this case one has the following representation: introducing local relaxation rates (longitudinal and transversal)

\begin{equation}\label{rates}
  \gamma_{\rm L}(t) := \gamma_+(t) + \gamma_-(t) \ , \ \ \ \gamma_{\rm T}(t) := \frac{\gamma_+(t) + \gamma_-(t)}{2} + 2 \gamma_z(t)  ,
\end{equation}
and

\begin{equation}\label{}
  \Gamma_{\rm L}(t) = \int_0^t \gamma_{\rm L}(\tau) d\tau \ , \ \ \  \Gamma_{\rm T}(t) = \int_0^t \gamma_{\rm T}(\tau) d\tau \ ,
\end{equation}
one finds \cite{Sabrina-NJP}
\begin{equation}\label{}
  B(t) = \exp(- \Gamma_{\rm L}(t)/2 ) \ ,  \ \ \  C(t) = \exp(-\Gamma_{\rm T}(t))   .
\end{equation}

 Actually, for any qubit dynamical semigroup a set of three relaxation rates $\{G_1,G_2,G_3\}$ satisfy the following relation \cite{Gen3}

\begin{equation}\label{G=2}
  G_k \leq \frac 12 (G_1+G_2+G_3) = \frac 12 G_{\rm total} .
\end{equation}
Recall, that $G_k = - {\rm Re}\,\ell_k$ and $\ell_k$ are (in general complex) eigenvalues of the qubit generator, i.e. $\mathcal{L}(X_k) = \ell_k X_k$. Interestingly, in a recent paper \cite{Gen} (cf. also \cite{Gen2}) it was conjectured that the above relation can be generalized for any dynamical semigroup of $d$-level quantum system as follows

\begin{equation}\label{}
  G_k \leq \frac 1d \sum_{m=1}^{d^2-1} G_m = \frac 1d G_{\rm total} .
\end{equation}
This relation is satisfied for several important classes of Markovian semigroups \cite{Gen}. Moreover, it is strongly supported by numerical analysis. Now, if the generator (\ref{L}) is time-dependent and all three transition rates $\gamma_\pm(t) \geq 0$ and $\gamma_z(t) \geq 0$, then  the local time-dependent relaxation rates

\begin{equation}\label{}
  G_1(t) = G_2(t) = \gamma_{\rm T}(t) \ , \ \ \ G_3(t) = \gamma_{\rm L}(t)
\end{equation}
satisfy (\ref{G=2}) for any $t \geq 0$, that is,

\begin{equation}\label{!!}
 \gamma_{\rm T}(t) \leq \frac 12  \gamma_{\rm total}(t) \ , \ \ \  \gamma_{\rm L}(t) \leq \frac 12  \gamma_{\rm total}(t) ,
\end{equation}
where the total rate reads

\begin{equation}\label{}
   \gamma_{\rm total}(t)  = 2( 	\gamma_+(t) + \gamma_-(t) + 2 \gamma_z(t) ) .
\end{equation}
Conditions (\ref{!!}) are necessary for CP-divisibility of he corresponding dynamical map. Therefore, for the non-Markovian evolution the validity of (\ref{!!}) is not guaranteed. Note, that conditions (\ref{!!}) imply

\begin{equation}\label{}
 \gamma_{\rm L}(t) = \gamma_+(t) + \gamma_-(t) \geq 0 \ , \ \ \ \gamma_z(t) \geq 0 ,
\end{equation}
and hence (\ref{!!}) guarantee that the evolution satisfies BLP condition  (\ref{BLP1}). Interestingly, $ \gamma_{\rm total}(t) \geq 0$ is equivalent to the geometric criterion (\ref{GEO1}).  However, as we shall see in general conditions (\ref{!!}) are violated (cf. Section \ref{EXP}).

\section{Exponential profile: a case study}    \label{EXP}

 {We illustrate the behaviour of the system and the problem of non-Markovianity of its evolution for the exponential profile
\begin{equation}
\xi({t}) =\begin{cases}
0& \mathrm{ for}\;\; t<0\\
\sqrt{\Gamma_p}e^{-\Gamma_p t/2} & \mathrm{ for}\;\; t\geq 0 .
\end{cases}
\end{equation}
The wave-packet of such shape can be experimentally produce by controlled emission \cite{Forn17}. We perform the analysis introducing the positive parameter $\alpha$ and writing $\Gamma_{p}$ as a fraction of the total coupling constant $\Gamma$ i.e. $\Gamma_p = \alpha \Gamma $. From the general formulae (\ref{A(t)})-(\ref{C(t)}), we obtain then
\begin{equation}\label{A_out_of_resonce}
A(t)=\frac{4 \kappa\alpha \Gamma^2}{(1-\alpha)^2\Gamma^2+4\Delta_{0}^2} \left(e^{-\Gamma t}-2e^{-\frac{(1+\alpha)\Gamma t}{2}}\cos  \Delta_{0} t+e^{-\alpha\Gamma t}\right)
\end{equation}
\begin{eqnarray}\label{B_out_of_resonce}
B(t)&=&e^{-\Gamma t} \left\{1-\frac{8\kappa\Gamma}{(1+\alpha)^2\Gamma^2+4\Delta_{0}^2}
\Bigg[(1+\alpha)\Gamma\left(e^{-\alpha\Gamma t}-1\right)\nonumber
\right.\\
&&\left.+\frac{2\alpha\Gamma}{(1-\alpha)^2\Gamma^2+4\Delta_{0}^2}\bigg[\left((1-\alpha^2)\Gamma^2-4\Delta_{0}^2\right)
\left(e^{\frac{(1-\alpha)\Gamma t}{2}}\cos \Delta_{0} t-1\right)+4\Delta_{0}\Gamma e^{\frac{(1-\alpha)\Gamma t}{2}}\sin \Delta_{0} t\bigg]\Bigg]\right\}
\end{eqnarray}
\begin{equation}\label{C_out_of_resonce}
C(t)=e^{-(i\Delta_{0}+\frac{\Gamma }{2})t}
\left\{1-	\frac{4\kappa\Gamma}{(\Gamma-2i\Delta_{0})^2-\alpha^2\Gamma^2} \left[(-2i\Delta_{0} +(1+\alpha)\Gamma)(1-e^{-\alpha\Gamma t})+2\alpha\Gamma\left(e^{\left(i\Delta_{0}-\frac{(1+\alpha)\Gamma}{2}\right)t}-1\right)\right]\right\} .
\end{equation}

Let us note that the formula for $A(t)$ allows one to compute the maximal excitation probability for an exponential profile.

\begin{Proposition}\label{PRO-4} The maximal excitation probability for an exponential profile reads
	\begin{equation}\label{Pmax}
	P_e^{\rm max} = \frac{4\kappa}{e^2}.
	\end{equation}
	and is realized for $\alpha=1$ and $\Delta_{0}=0$ at $t = 2/\Gamma$.
\end{Proposition}
For the proof cf. Appendix \ref{max_exp}. Recently, this result for $\kappa=0.5$ was derived via different methods in \cite{Branczyk} (it was found numerically in \cite{WMSS11}). Clearly, the fewer $\kappa$ one takes, the smaller excitation one obtains. We mention it here because it indicates the values of the parameters for which we observe the maximum influence of the single-photon field on the evolution of the atom. From the physical point of view the cases when $\alpha\ll 1 $ or $\alpha\gg 1$, or when the central frequency of the field is far from the resonance they are rather not interesting. Briefly, if $\alpha \ll 1$, then the interaction between the atom and the single-photon field is too weak to drive the system. If $\alpha \gg 1$, then the mean time of interaction, given by $\Gamma_{p}^{-1}=\alpha^{-1}\Gamma^{-1}$, is too short to change the state of the system. The evolution goes then to the semigroup. One can notice it considering the respective limits of formulae (\ref{A_out_of_resonce})-(\ref{C_out_of_resonce}).

\subsection{Resonant case $\Delta_0 = 0$}

First, we carefully analyse the resonant case. Then, if $\alpha\neq 1$ we get
\begin{equation}\label{A_resonant}
A(t)=\frac{4 \kappa\alpha}{(1-\alpha)^2} \left(e^{-\frac{\Gamma t}{2}}-e^{-\frac{\alpha\Gamma t}{2}}\right)^2
\end{equation}

\begin{eqnarray}\label{B_resonant}
B(t)&=&\frac{e^{-(1+\alpha)\Gamma t}}{\alpha^2-1} \left[16\kappa \alpha e^{\frac{(1+\alpha)\Gamma t}{2}}+[\alpha^2-8\kappa(1+\alpha)-1]e^{\alpha \Gamma t}-8\kappa(\alpha-1)\right]
\end{eqnarray}

\begin{equation}\label{C_resonant}
C(t)=\frac{e^{-(1+\alpha)\Gamma t}}{\alpha^2-1}
\left[(1-\alpha)(4\kappa-\alpha-1)e^{\left(\alpha+\frac{1}{2}\right)\Gamma t}-4\kappa(1+\alpha)e^{\frac{\Gamma t}{2}}+8\kappa \alpha e^{\frac{\alpha\Gamma t}{2}}\right].
\end{equation}
And for $\alpha=1$, we have the formulae
\begin{equation}\label{A_resonant_1}
A(t)= \kappa\Gamma^2 t^2e^{-\Gamma t},
\end{equation}

\begin{equation}\label{B_resonant_1}
B(t)=e^{-\Gamma t} \left\{1-	4 \kappa\Gamma\left[t+\frac{1}{\Gamma}(e^{-\Gamma t}-1)\right]\right\},
\end{equation}

\begin{equation}\label{C_resonant_1}
C(t)=e^{-\frac{\Gamma t}{2}}
\left\{1-2\kappa\Gamma\left[-te^{-\Gamma t}-\frac{1}{\Gamma}\left(e^{-\Gamma t}-1\right)\right]\right\}.
\end{equation}
Now, using (\ref{gg}) and (\ref{gz}) one computes the expressions for $\gamma_{+}$, $\gamma_{-}$, and $\gamma_{z}$ (see Appendix \ref{Appengammas}). Let us notice that even for a relatively simple exponential pulse in resonance the formulae are quite involved. However, they do allow to provide the analysis of memory effects of the corresponding evolution. For a resonant case one can easily discriminate between invertible and non-invertible dynamical maps.

\begin{Proposition}\label{PRO-5} The evolution is invertible iff $\alpha \geq 8 \kappa + 1$.
\end{Proposition}
For the proof cf. Appendix \ref{AppenD}. Thus, for a given $\kappa$, we can always indicate the range of $\alpha$ producing an invertible evolution.

\begin{Corollary} In the regular case, i.e. $\alpha \geq 8 \kappa + 1$, the evolution satisfies BLP condition, i.e. there is no information backflow.
\end{Corollary}
Numerical analysis also shows that in the resonant case P-divisibility and BLP condition coincide. However, we were not able to provide analytical proof. The two scenarios: for invertible and non-invertible maps for $\kappa=1$ are illustrated in  Fig. \ref{Fig-BB}. For the regular case, defined then by $\alpha\geq 9$, we observe the monotonic decreasing functions $B(t)$ and $C(t)$ which remain positive and approach zero values for large time. Note that  for non-invertible maps, whenever $B(t_*)=0$, then $B(t)$ stays negative for all $t > t_*$ and asymptotically approaches zero. The same we observe for $C(t)$. Thus in the second scenario, first we see a decrease of coherence, then we deal with increase of coherence, and finally we observe the effect of decoherence. The case of non-invertible map is illustrated by $\alpha=1$ giving the maximal excitation of the atom. In Fig. \ref{Fig-BB}  the probability of excitation, $P_{e}(t)$, for a chosen initial state and the function $A(t)$ are also depicted. Of course, $P_{e}(t)$ as well as $A(t)$ are non-monotonic functions of time in the both, regular and singular, cases. Note that for the regular case we deal with the eternally non-Markovian evolution i.e  $\gamma_{z}(t)<0$ for all $t>0$ as depicted in Fig. \ref{Fig-ggg1}. A singularity of the generator $\mathcal{L}_t$, displaying for $\kappa=1$  whenever $\alpha<9$, manifests in the singularities of all rates which are shown for $\alpha=1$ in Fig. \ref{Fig-ggg1}. It is interesting (see Fig. \ref{Fig-ggg1}) that $\gamma_{+}(t)$ and $\gamma_{z}(t)$ have always opposite signs:
\begin{equation}
\forall t\geq 0\;\;\;\;\gamma_{+}(t)\gamma_{z}(t)\leq 0.
\end{equation}
In general, one sees that at least one of the three rates is negative at a given time.

\begin{figure}[h]
\includegraphics[width=6cm]{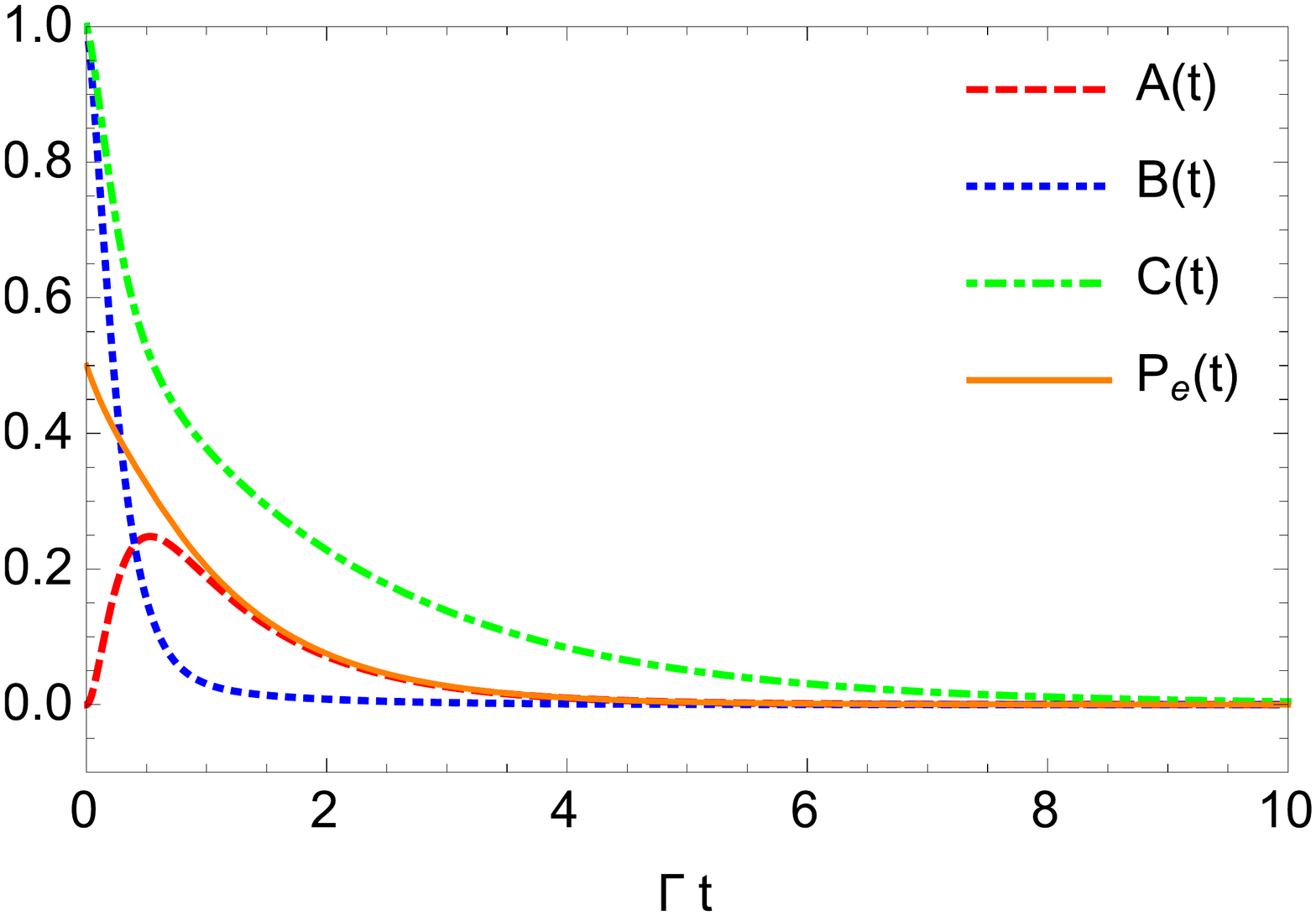}
\includegraphics[width=6cm]{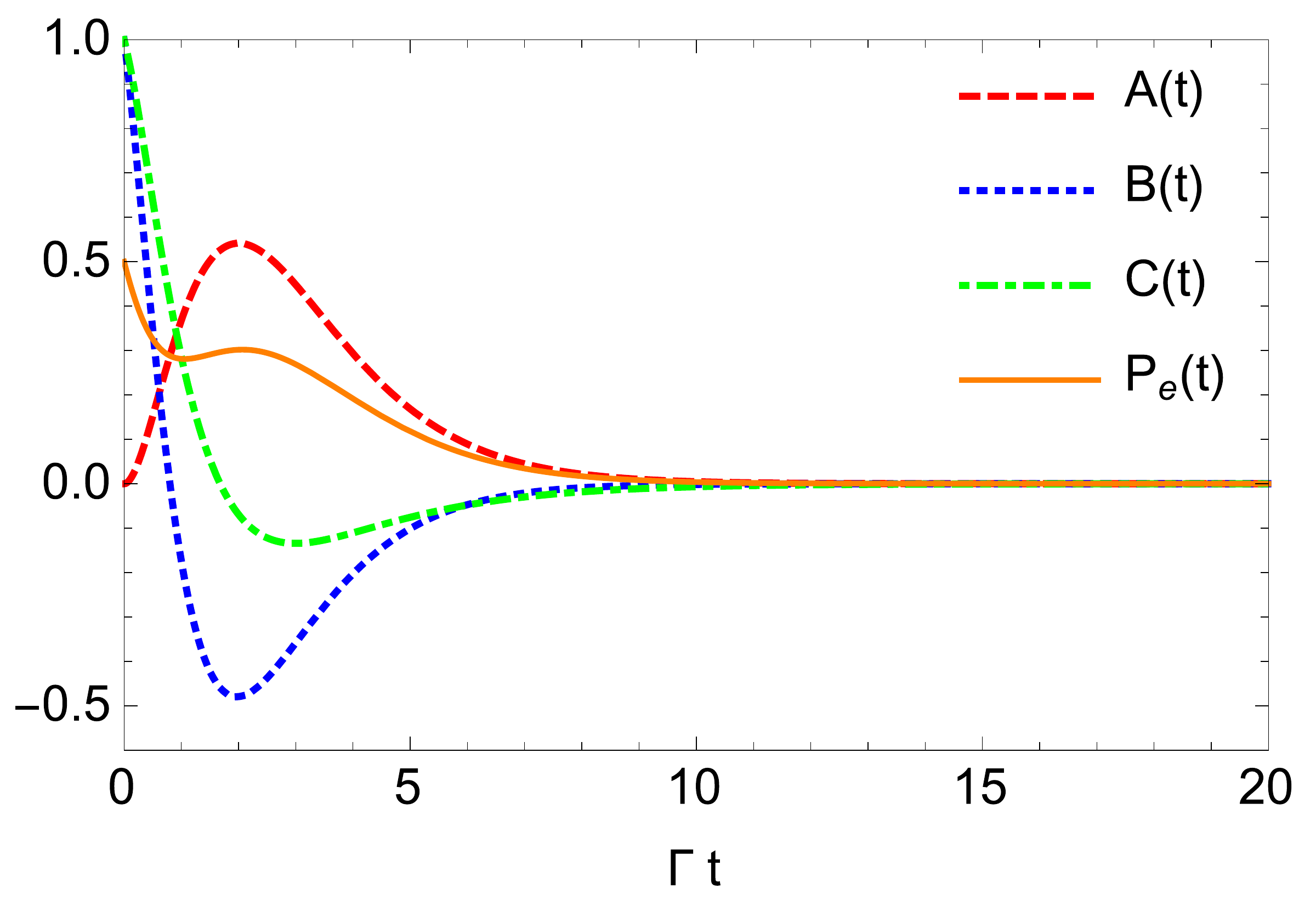}
	\caption{The plots of $A(t)$, $B(t)$, $C(t)$, and $P_{e}(t)$ for $\kappa=1$, $\Delta_{0}=0$ and $P_{e}(0)=0.5$. Left: regular case $\alpha=9.5$. Right: singular case $\alpha =1$.}\label{Fig-BB}
\end{figure}

\begin{figure}[h]
\includegraphics[width=6cm]{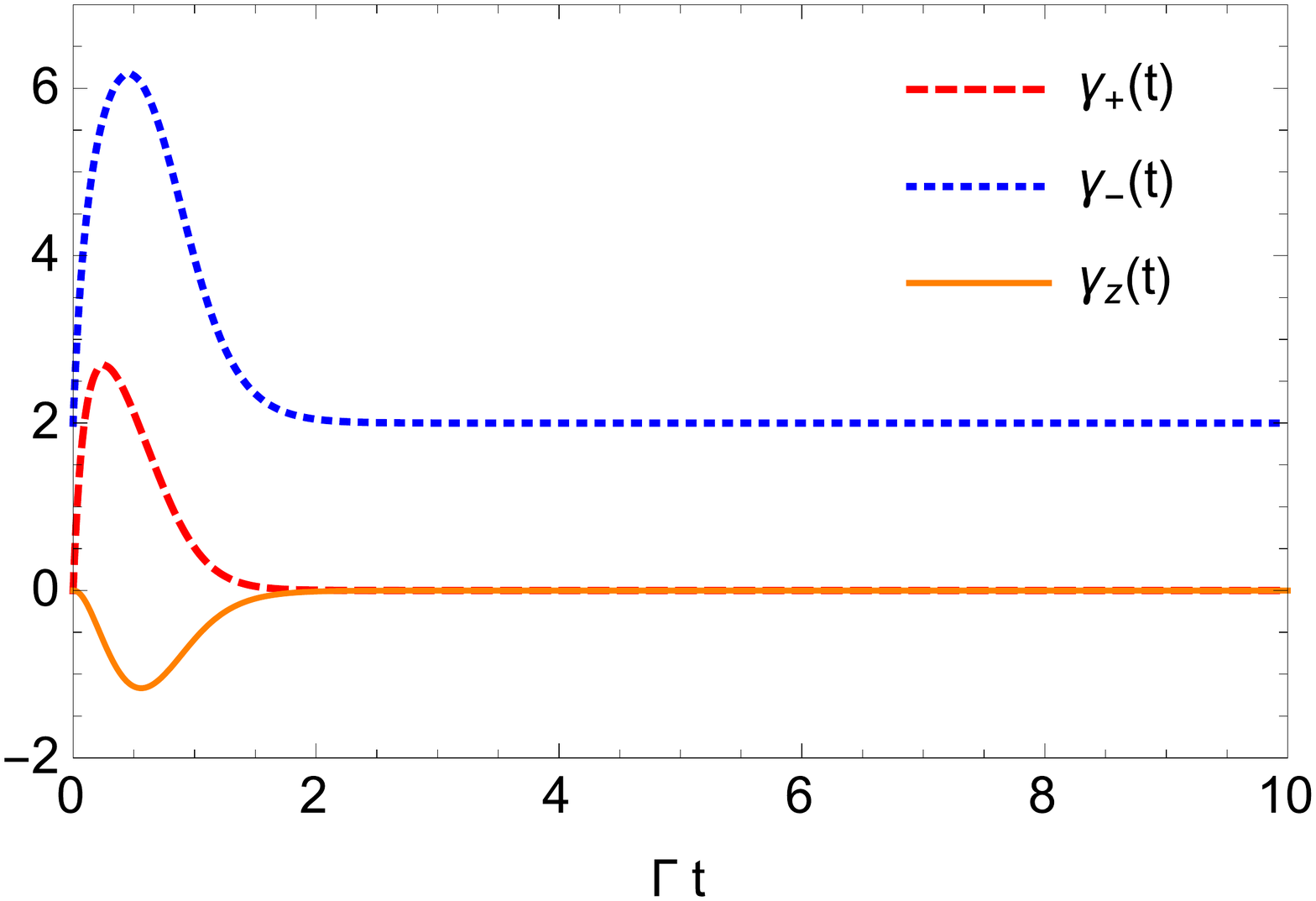} \includegraphics[width=6cm]{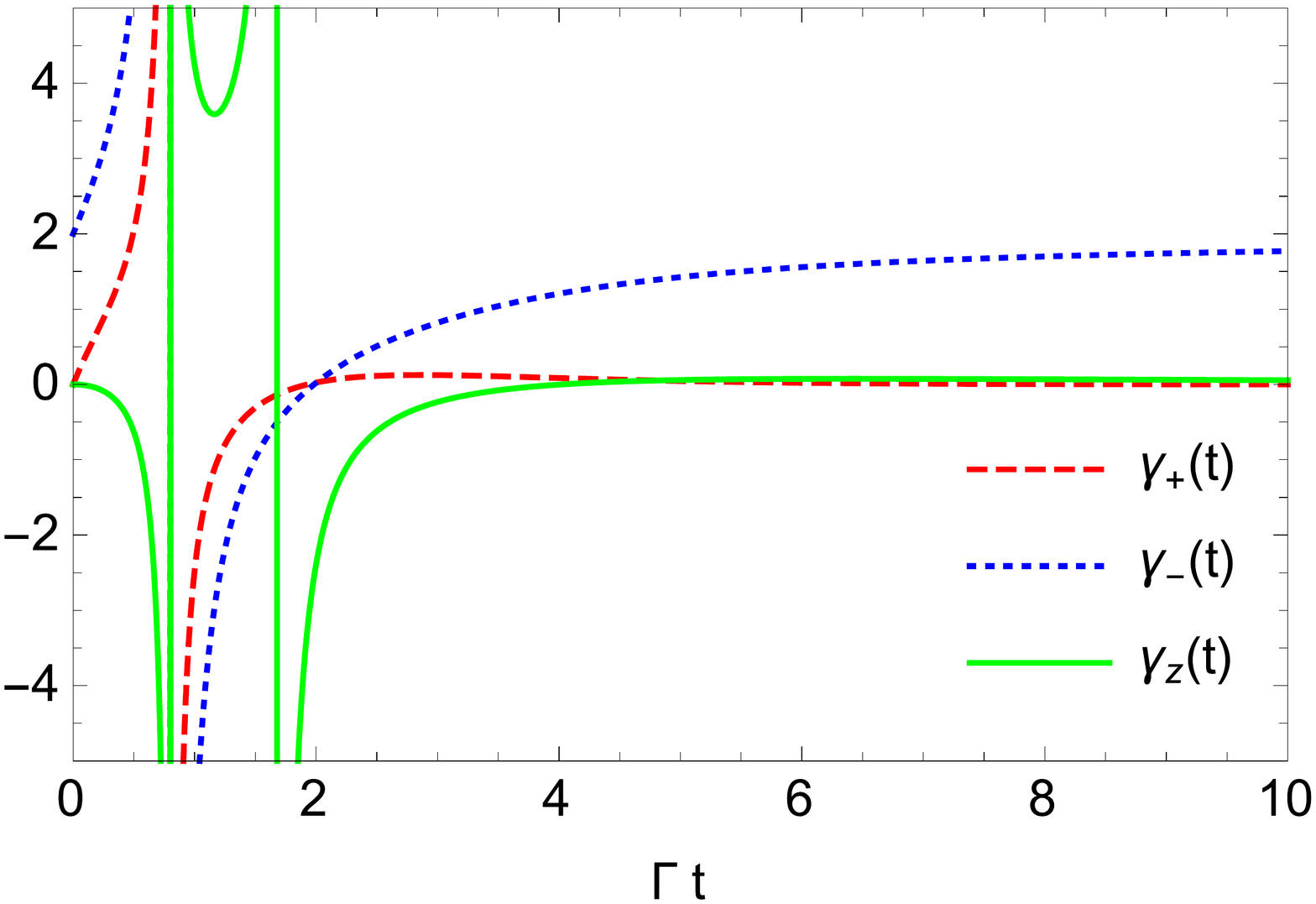}
\caption{The plots of $\gamma_{+}(t)$, $\gamma_{-}(t)$, and $\gamma_{z}(t)$ for $\Gamma=1$, $\kappa=1$, $\Delta_{0}=0$. Left: regular case $\alpha=9.5$. Right: singular case $\alpha=1$.} \label{Fig-ggg1}
\end{figure}

\begin{Proposition} One has the following asymptotic behaviour  for $\gamma_\pm(t)$ and $\gamma_z(t)$:

	\begin{itemize}
		\item for $\alpha\geq 1$
		
		\begin{equation}\label{}
		\lim_{t\to \infty} \gamma_+(t) =  \lim_{t\to \infty} \gamma_z(t) = 0 \ , \ \ \  \lim_{t\to \infty} \gamma_-(t) =  2\Gamma
		\end{equation}
		
		\item for $\alpha\in (0,1)$

		\begin{equation}\label{}
		\lim_{t\to \infty} \gamma_+(t) =  0 \ , \ \ \ \lim_{t\to \infty} \gamma_z(t) = \frac 14 \Gamma (1-\alpha)  \ , \ \ \  \lim_{t\to \infty} \gamma_-(t) =  \Gamma (1+\alpha).
		\end{equation}
	\end{itemize}
\end{Proposition}
It implies that asymptotically one has

\begin{itemize}
	\item for $\alpha \geq  1$
	
	\begin{equation}\label{Ia}
	\mathcal{L}_t(\rho) \ \to \   {\Gamma} \left( \sigma_- \rho \sigma_+ - \frac 12 \{ \sigma_+ \sigma_-,\rho\} \right)
	\end{equation}
	\item for $\alpha\in(0, 1)$
	
	\begin{equation}\label{IIa}
	\mathcal{L}_t(\rho)  \ \to \    \frac{\Gamma(1+\alpha)}{2} \left( \sigma_- \rho \sigma_+ - \frac 12 \{ \sigma_+ \sigma_-,\rho\} \right) + \frac{\Gamma(1-\alpha)}{8} ( \sigma_z\rho\sigma_z -\rho) .
	\end{equation}
\end{itemize}

Let us note that the asymptotic evolution of coherence governed by (\ref{Ia}) and (\ref{IIa}) is the same. For details cf. Appendix \ref{Appengammas}.

}

\subsection{Off-resonant case $\Delta_0 \neq 0$}

Now, the formulae for $A(t)$, $B(t)$, and $C(t)$ are much more involved and we were not able to find analytical condition for $\{\Delta_0,\Gamma,\Gamma_p\}$ which guaranties invertible dynamical map $\Lambda_t$. { We observe  that contrary to the resonant case $B(t)$ may vanish at several moments of time which makes the generator even more singular than in the resonant case. Figure \ref{multi} shows that increasing $\Delta_0$ from `0' (resonant case) to $\Delta_{0}=1.5$ one creates additional singular points for $\kappa=1$ and $\alpha=1.5$. However, if $\Delta_0=2.5$ all singularities are removed.
Second interesting observation is related to indicators of non-Markovianity. In the resonant case if the generator is free from singularities (i.e. $\alpha \geq 8 \kappa + 1$) the evolution satisfies BLP condition and hence does not display information backflow. It is no longer true if $\Delta_0 \neq 0$. Figure \ref{yes-no} shows that even if the generator is free from singularities the validity of BLP condition still depends for a given $\kappa$ on the parameters  $\{\Delta_0,\alpha\}$. Since $\gamma_+(t)$ might be negative, it is clear, that even if the dynamics satisfies BLP condition it needs not be P-divisible (cf. Theorem \ref{TH}). Therefore, in the off-resonant case these two indicators, that is, P-divisibility and BLP condition do not coincide. This observation is quite interesting since in majority of studied examples these two notions coincide. It simply means that for off-resonant evolution the monotonicity condition

\begin{equation}\label{}
\frac{d}{dt}\| \Lambda_t(X) \|_1 \leq 0,
\end{equation}
is violated even if it is satisfied for all traceless Hermitian operators (i.e. $X \sim \rho_1 - \rho_2$). This analysis shows that no information backflow does not guarantee the existence of positive trace-preserving propagators $V_{t,s}$. Actually, one may violate not only BLP condition but much weaker condition based on the geometric indicator \cite{LPP13} which measures the volume of accessible states. One finds that if the evolution is Markovian then

\begin{equation}\label{}
\frac{d}{dt} |{\rm det}\Lambda_t| =  \frac{d}{dt} |\lambda_1(t)\lambda_2(t)\lambda_3(t)|\, \leq\,  0 ,
\end{equation}
where $\lambda_k(t)$ denote eigenvalues of the dynamical map ($k=1,2,3$). This condition is equivalent to (\ref{GEO1}) \cite{Sabrina-NJP}, which is much weaker than BLP condition.  The geometric and the BLP criteria of non-Markovianity are shown in Fig. \ref{Fig-geo2}. One sees that for parameters indicated in the left panel of Figure \ref{Fig-geo2} this condition is violated proving that evolutions display strong memory effects. Finally, let us note that Fig. \ref{yes-no} shows that for the chosen values of parameters the rates oscillate with decreasing amplitudes around their asymptotic values, namely $\gamma_{+}(t)$ and $\gamma_{z}(t)$ oscillate around zero, and $\gamma_{-}(t)$ around $2\Gamma$. Interestingly, whenever $\gamma_{+}(t)$ is zero $\gamma_{z}(t)$ is also equal to zero, and $\gamma_{-}(t)$ is equal to $2\Gamma$, so for these moments $\mathcal{L}_{t}=\mathcal{L}$.

\begin{figure}
\includegraphics[width=5.5cm]{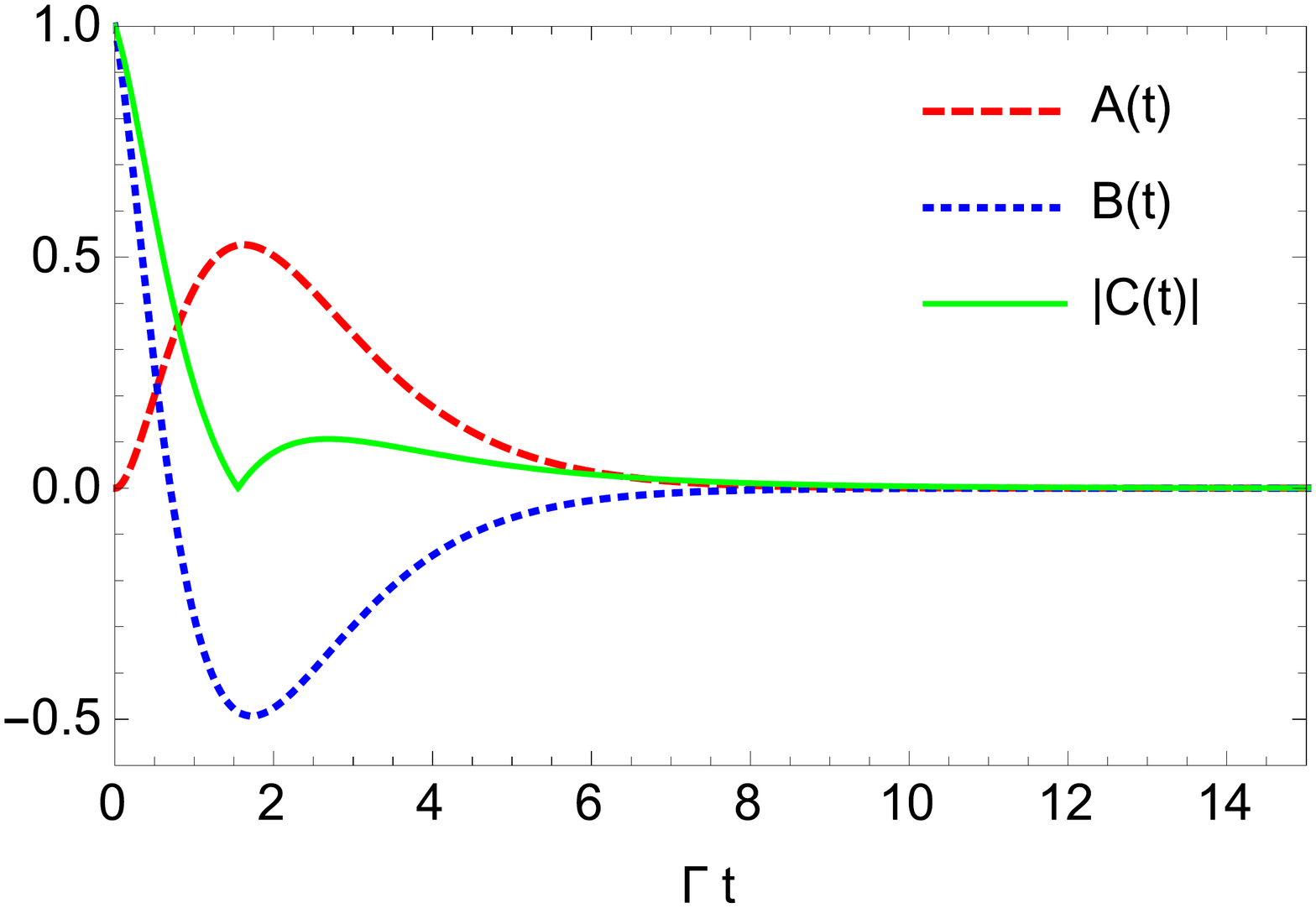}  \includegraphics[width=5.5cm]{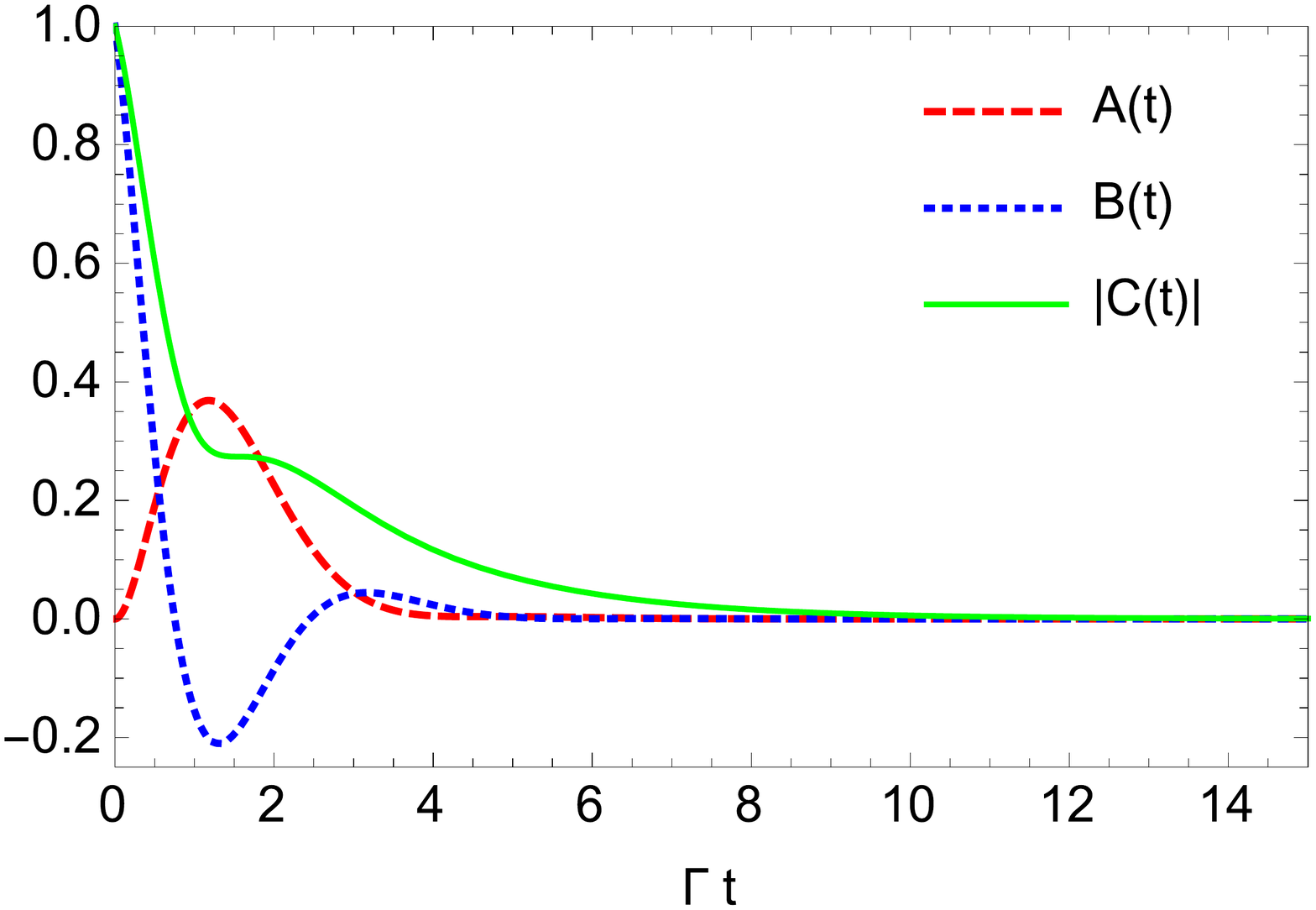}
\includegraphics[width=5.5cm]{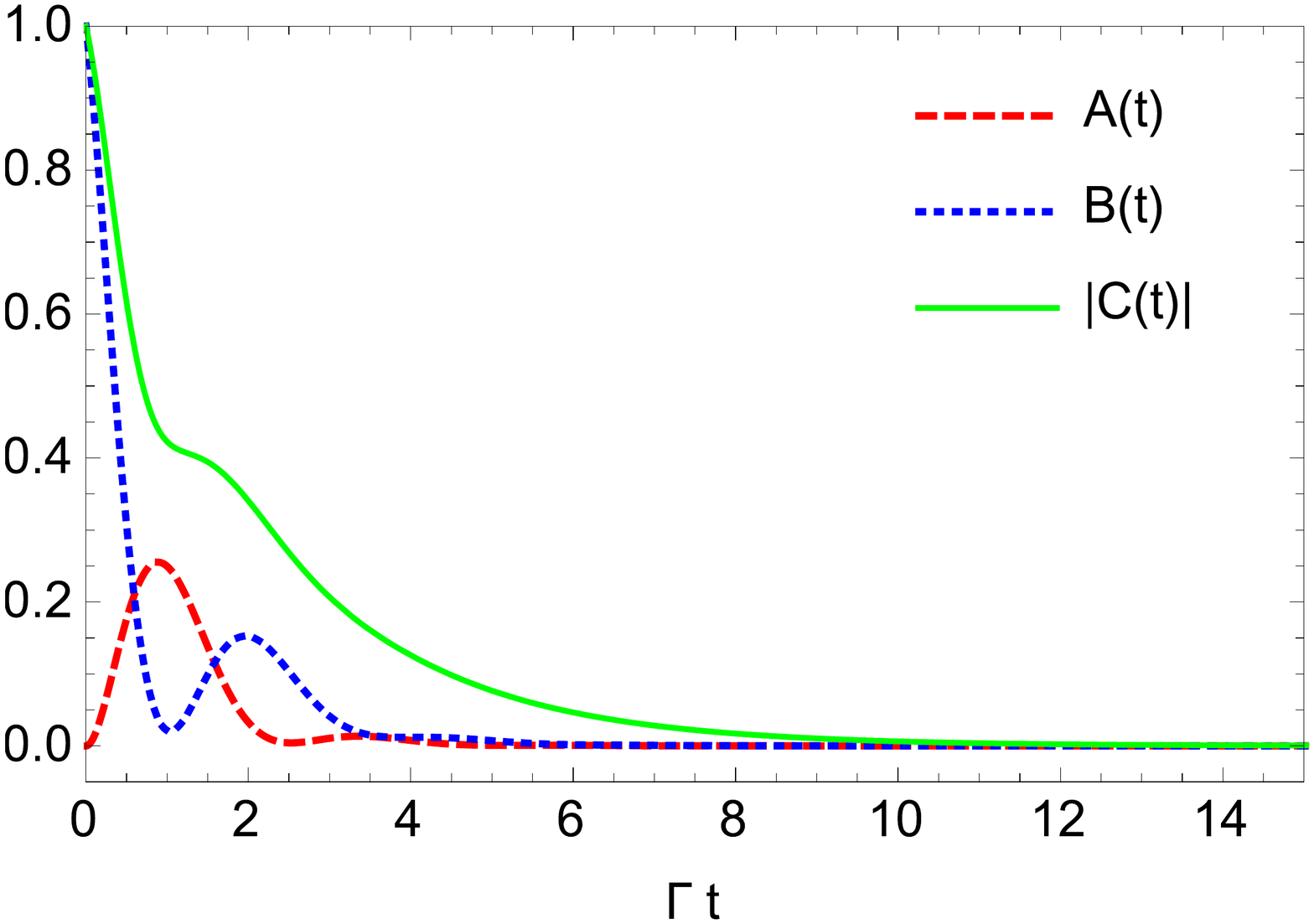}
\caption{The plots of $A(t)$, $B(t)$, and $|C(t)|$ for $\kappa=1$, $\Gamma=1$, and $\alpha=1.5$. Left: $\Delta_0=0$. Center: $\Delta_0=1.5$. Right: $\Delta_0=2.5$.} \label{multi}
\end{figure}

\begin{figure}
	\includegraphics[width=6cm]{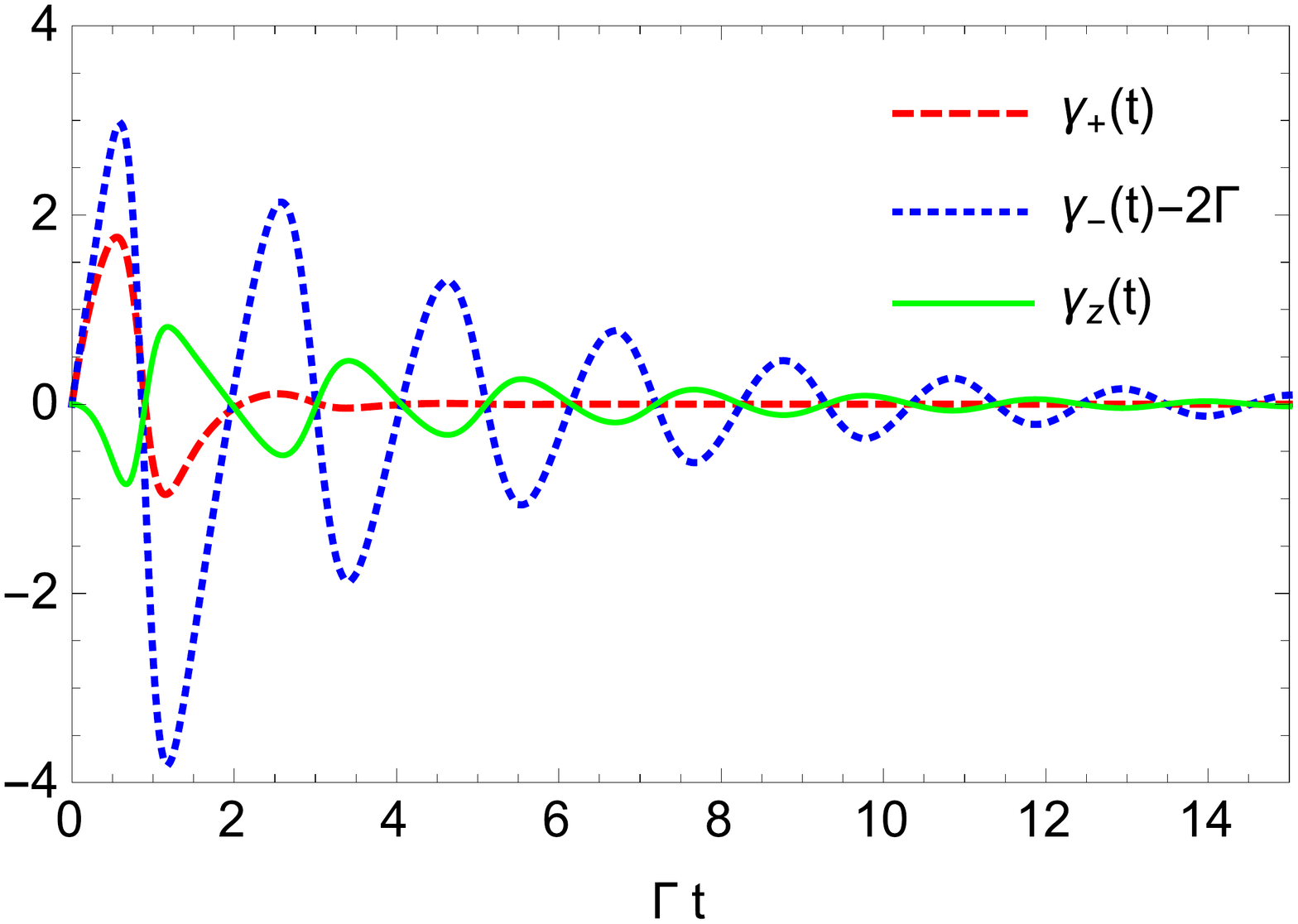}   \includegraphics[width=6cm]{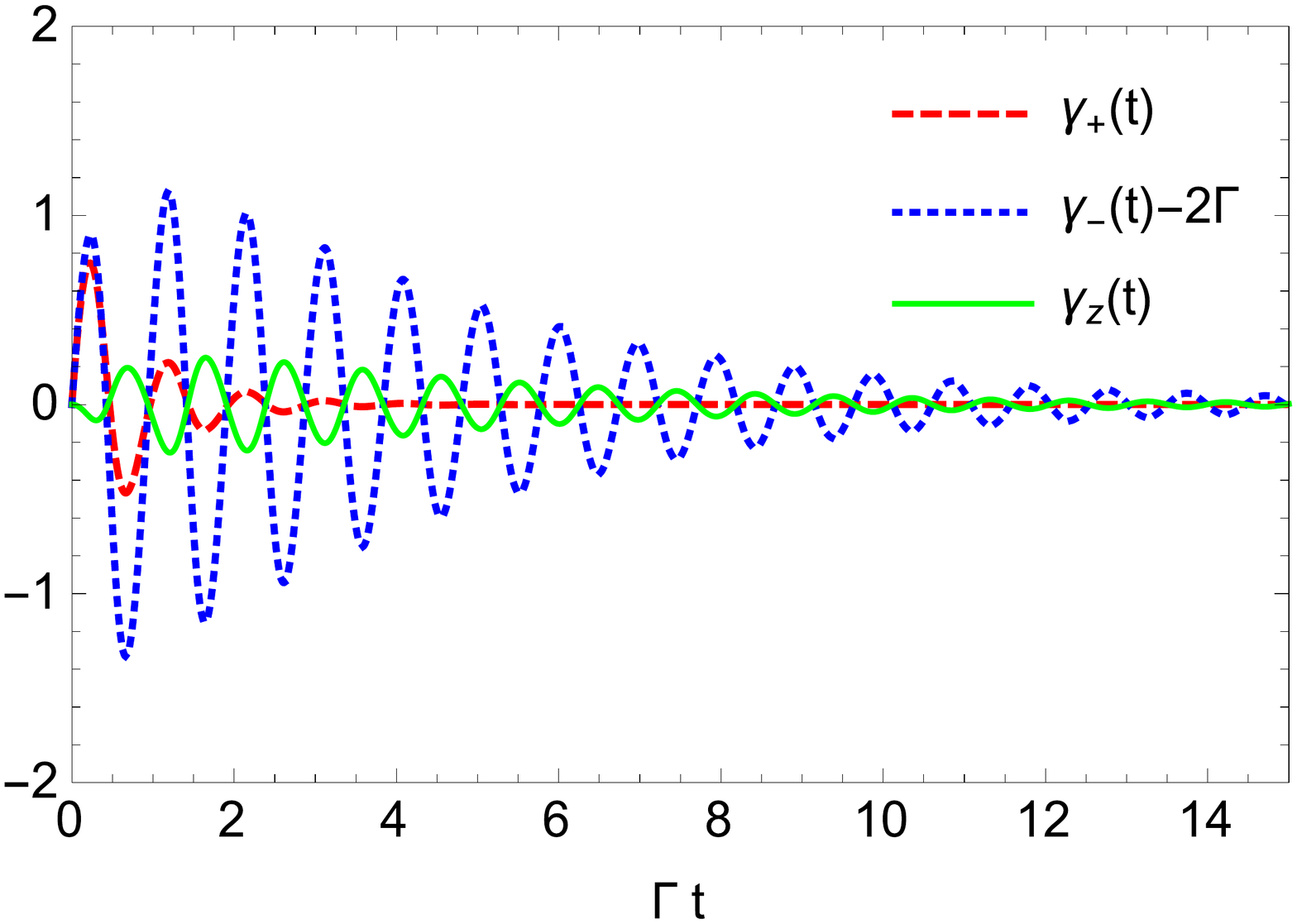}
	\caption{The plots of $\gamma_{+}(t)$, $\gamma_{-}(t)-2\Gamma$, and $\gamma_{z}(t)$ for $\kappa=1$, $\Gamma=1$, and $\alpha=1.5$. Left: $\Delta_0=3.0$ --- BLP condition is violated. Right: $\Delta_0=6.5$ --- BLP condition is satisfied.} \label{yes-no}
\end{figure}

\begin{figure}[h]
		\includegraphics[width=6cm]{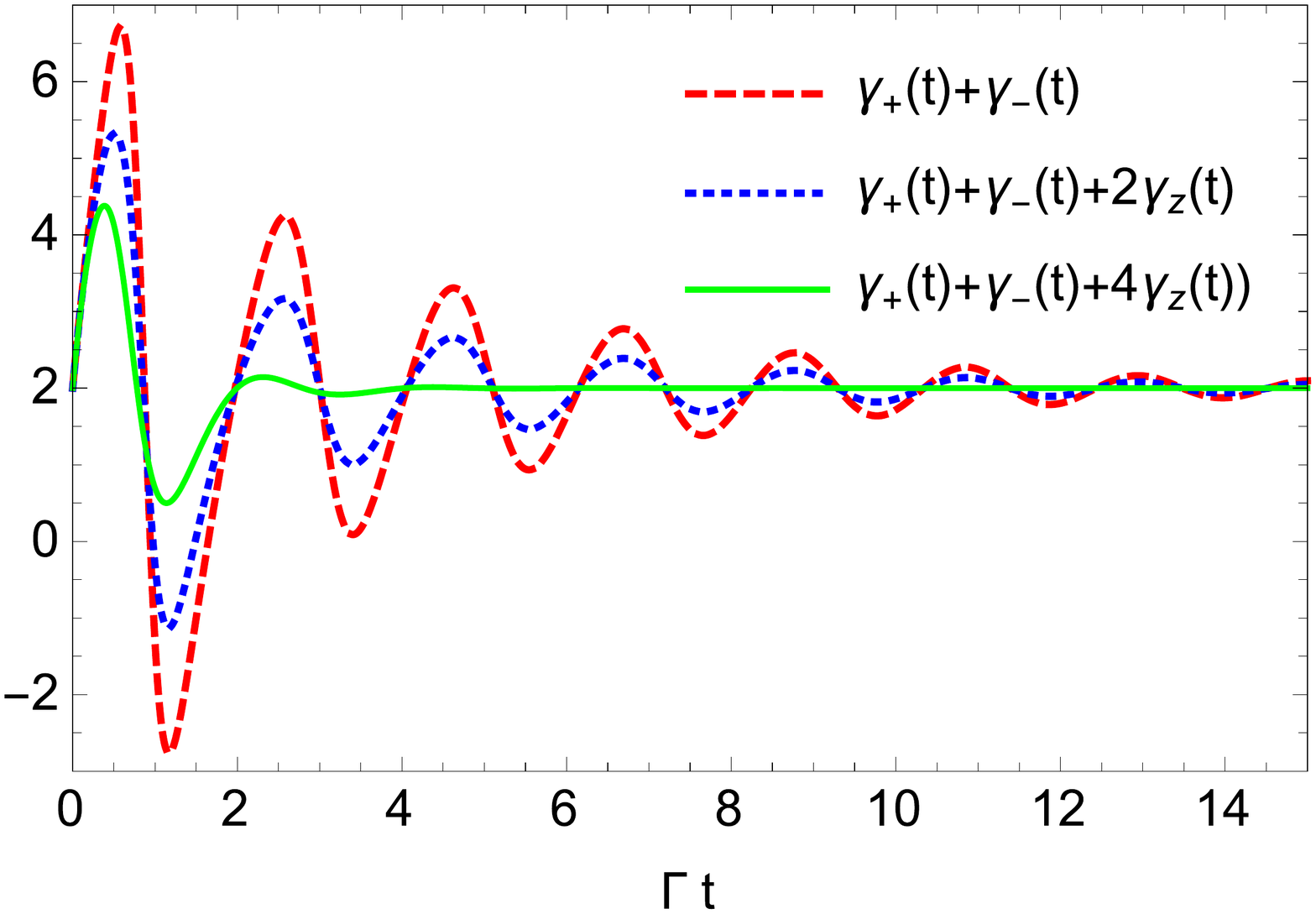} \includegraphics[width=6cm]{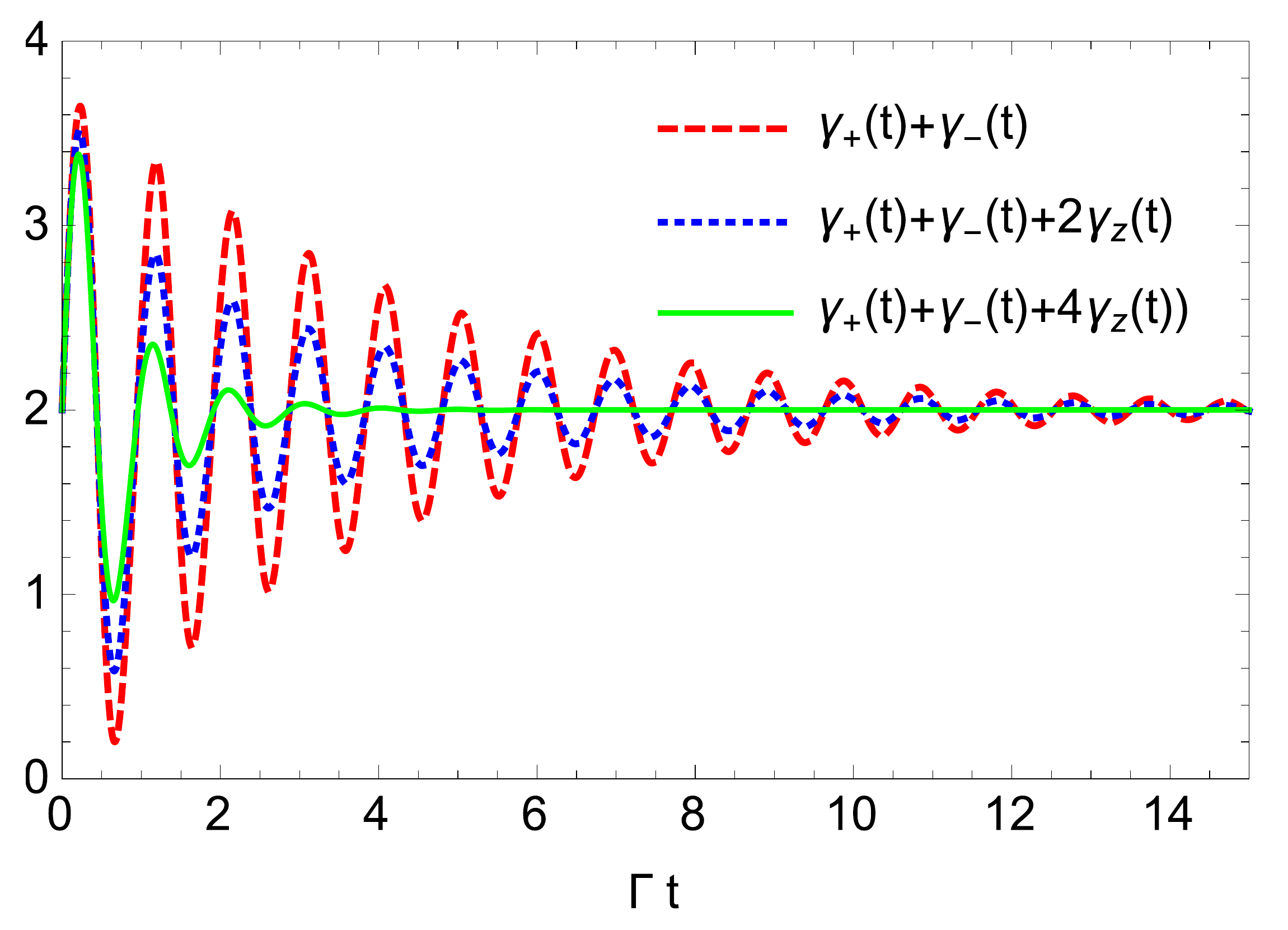}
		\caption{The plots of $\gamma_{+}(t)+\gamma_{-}(t)$,  $\gamma_{+}(t)+\gamma_{-}(t)+2\gamma_{z}(t)$, and $\gamma_{+}(t)+\gamma_{-}(t)+4\gamma_{z}(t)$ for $\kappa=1$, $\Gamma=1$,  $\alpha=1.5$. Left: $\Delta_0=3.0$. Right: $\Delta_0=6.5$.} \label{Fig-geo2}
	\end{figure}

\section{Conclusion}\label{S-VIII}

We provided a detailed discussion of the evolution of a qubit interacting with single-photon wave packet. Clearly, this problem was already analyzed by several researchers (see e.g.  \cite{WMSS11,SAL10a,Francesco,Branczyk}).
In this paper it is shown that a set of hierarchical equations, already derived in \cite{GEPZ98,BCBC12,DSC17,DSC19,D20}, is equivalent to a single  time-local master equation for the qubit density operator. To show this equivalence we make use of the general form of solution to this set defining a dynamical map $\{\Lambda_t\}_{t\geq 0}$ for an arbitrary photon profile $\xi(t)$. The price one pays for this reduction is highly nontrivial structure of time-dependent rates in the time-local master equation. Interestingly, the corresponding time-local generator has a structure of phase covariant generator which has been already extensively studied  \cite{SKHD16,Sabrina-2016,Sabrina-NJP,Francesco,HSH19,Sergey-2020}. We provided the formulae for time dependent rates governing damping (cooling), heating and decoherence processes. These rates are fully controlled by the photon profile $\xi(t)$ which is a key ingredient of the entire analysis. It turn out that temporal correlations encoded into $\xi(t)$ give rise to all characteristic non-Markovian memory effects of the corresponding qubit evolution. Due to the interaction with a non-locally (in time) entangled field the system undergoes such phenomena as recoherence. But the recoherence is not universal, for some parameters the photon becomes only the source of decoherence.

Note that a dynamical map $\{\Lambda_t\}_{t\geq 0}$ representing the qubit evolution might be non-invertible for some time moments which implies the singularity of transition rates in the corresponding time-local master equation. The phenomenon of non-invertible maps were already considered in quantum optics  for example in \cite{Cresser1,Cresser2,SKHD16}. Similar observation for phase covariant dynamics was also recently reported in \cite{Francesco}. It should be stressed that being singular the time-local generator $\mathcal{L}_t$ gives rise to perfectly regular evolution \cite{D20} (represented by $\{\Lambda_t\}_{t\geq 0}$). Non-invertible dynamical maps were recently analyzed in \cite{PRL-2018} and \cite{CC19} in connection to divisibility (cf. also recent paper \cite{Jyrki-2021}). Divisibility which is equivalent to the existence of a propagator $V_{t,s}$ is always guaranteed for invertible maps. For maps which are not invertible  divisibility holds only under additional conditions \cite{PRL-2018}.

We provided the detailed analysis of the exponential profile. In particular we derived the formula for the maximal excitation probability. Such formula was also derived in a recent paper \cite{Branczyk} via different methods. For exponential profile we derived explicit formulae for all time dependent rates of the time-local generator. The key observation is that for the  exponential profile qubit evolution is never Markovian (i.e. CP-divisible). Interestingly, in the resonant case the dynamical map $\{ \Lambda_t\}_{t \geq 0}$ is either invertible or looses invertibility in at most two moments of time. However, whenever it is invertible the evolution satisfies BLP condition which guarantees no information backflow. Off-resonant scenario is much more intricate since the number of singular points for the time-local generator highly depends upon the detuning parameter $\Delta_0$. Still, one can observe some universal properties of qubit dynamics. In particular $\gamma_+(t) \gamma_z(t) \leq 0$ for all $t \geq 0$. It shows that for any $t$ one of the transition rate is strictly negative and hence the dynamics is eternally non-Markovian \cite{Erika}. Similar analysis was performed for a square pulse $\xi(t)$. The conclusions are very similar and they are not included in the manuscript.

Beyond exponential profile it is still possible to observe universal properties. We shown that in the resonant case assuming $\xi(t)\in \mathbb{R}_{+}$ the evolution satisfies BLP condition whenever it is invertible. Still the evolution is eternally non-Markovian ($\gamma_z(t) \leq 0$) but invertibility protects the system against the information backflow. This observation clearly shows that being invertible is not only a mathematical curiosity but implies direct physical consequences.

The problem of non-Markovian dynamics of the qubit interacting with a
single-photon field was studied in \cite{Francesco} by making use an
infinite waveguide in the real-space approach. Authors of
\cite{Francesco} analyzed an exponential profile in resonance and
$\kappa= \frac 12$. We consider  off resonant case as well and
allow for $\kappa \in [0,1]$. Moreover, we derive general properties for an arbitrary profile satisfying $\xi(t)\in \mathbb{R}_{+}$. In particular it is shown that in the resonant case whenever the dynamical map is invertible the evolution satisfies BLP condition meaning there is no information backflow. Exponential profile is just an example of this general property.

The system we consider provides a perfect platform to test and compare various aspects of non-Markovianity. Detailed analysis of the
exponential profiles shows that the evolution of the qubit is never
Markovian (CP-divisible). Moreover, it is eternally non-Markovian
\cite{Erika}, that is, one of the transition rate is always negative. We claim that the same holds for an arbitrary profile meaning that the interaction with a single-photon field always results in the eternally non-Markovian dynamics. In our opinion this is the first proper physical model displaying such non-trivial behavior. The analysis of non-Markovian  dynamics of the system interacting with single-photon state is of particular importance for effective  control of
quantum systems and  the  storage and retrieving of quantum information. It would be interesting using similar theoretical methods to analyze the memory effects of a qubit interacting with $N$-photon field (the corresponding hierarchical equation were already derived in \cite{DSC19}).

\begin{acknowledgements}   This research was supported by the National Science Centre project 2018/30/A/ST2/00837.
\end{acknowledgements}

\appendix

\section{Solution of hierchical equations}\label{AppenA}

The solution to (\ref{mast2})-(\ref{mast4}) reads as
\begin{eqnarray}
{\varrho}^{01}(t)=\left( \begin{array}{ccc}
E(t)\rho_{eg}(0) & &F(t)  P_{e}(0)+G(t)\\
0 & & -E(t)\varrho_{eg}(0)
\end{array} \right)
\end{eqnarray}
\begin{eqnarray}
{\varrho}^{00}(t)&= \left(\begin{array}{ccc}
1-P_{e}(0)e^{-\Gamma t} & & \varrho_{ge}(0)e^{-\left(i\Delta_{0}+\frac{\Gamma}{2}\right) t} \\
\varrho_{eg}(0)e^{\left(i\Delta_{0}-\frac{\Gamma}{2}\right)t}
& & P_{e}(0)e^{-\Gamma t}
\end{array} \right)
\end{eqnarray}
\begin{equation}
E(t)=\sqrt{\kappa\Gamma}e^{- \Gamma t} \int_{0}^{t}ds \xi^{\ast}(s)e^{\left(i\Delta_{0}+\frac{\Gamma}{2}\right)s},
\end{equation}
\begin{equation}
F(t)=2\sqrt{\kappa\Gamma}e^{-\left(i\Delta_{0}+\frac{\Gamma}{2}\right)t} \int_{0}^{t}ds\xi^{\ast}(s)e^{\left(i\Delta_{0}-\frac{\Gamma}{2}\right) s},
\end{equation}
\begin{equation}
G(t)=-\sqrt{\kappa\Gamma}\int_{0}^{t}ds\xi^{\ast}(s)e^{\left(i\Delta_{0}+\frac{\Gamma}{2}\right)(s-t)},
\end{equation}
and $\varrho^{10}(t)=\left(\varrho^{01}(t)\right)^{\dagger}$.

\section{Proof of Proposition \ref{PRO-1}}\label{AppenB}

The proof is very simple: let $\tilde{\xi}(t)=e^{-i\omega_{c}t}\xi(t)\in\mathbb{C}$. Assuming  $\rho(0)=|g\rangle \langle g|$ one obtains

\begin{equation}\label{}
P_e(t)  = \kappa\Gamma e^{-\Gamma t} \left|\int_{0}^{t}ds \tilde{\xi}(s)e^{\left(i\omega_{0}+\frac{\Gamma}{2}\right) s}\right|^2.
\end{equation}
To maximize this expression one has to maximize
\begin{equation}\label{}
\left|\int_{0}^{t}ds \tilde{\xi}(s)e^{\left(i\omega_{0}+\frac{\Gamma}{2}\right) s}\right|^2 = |\langle\tilde{\xi}|f\rangle_t |^2
\end{equation}
where the function $f(s) =  e^{\left(-i\omega_{0}+\frac{\Gamma}{2}\right) s}$, and we introduce an inner product
\begin{equation}\label{}
\langle\tilde{\xi}|f\rangle_t := \int_0^t \tilde{\xi}(s) f^{\ast}(s) ds .
\end{equation}
It is evident that it is maximized for the profile $\tilde{\xi}(s)$ which is parallel to $f(s)$, that is, $\tilde{\xi}(s) = c(t) f(s)$, where $c(t)$ is a constant (depending on a fixed time $t$), and
\begin{equation}\label{}
\langle \tilde{\xi}|\tilde{\xi}\rangle_t = 1 .
\end{equation}
This normalization condition allows to calculate $c(t)$ leading to
\begin{equation}
\tilde{\xi}(\tau) = \sqrt{ \frac{\Gamma}{e^{\Gamma t}-1} } \, e^{\left(-i\omega_{0}+\frac{\Gamma}{2}\right) \tau}
\end{equation}
for $\tau \in[0,t]$, and $\tilde{\xi}(\tau)=0$ for $\tau > t$ and then to (\ref{A-max}). Clearly, the maximum excitation is realized at resonance  $\omega_{c}=\omega_{0}$, i.e. $\Delta_{0}=0$. \hfill $\Box$

\section{ Proof to Proposition \ref{PRO-2}}\label{AppenC}

Let us assume that
\begin{equation}\label{assumption}
\Delta_{0}=0, \ \ B(t)> 0 \ \ \mathrm{and} \ \ \xi(t)\in \mathbb{R}_{+} \ \ \mathrm{for \ \ all} \ \ t\geq 0.
\end{equation}
One can check that then
\begin{equation}
\dot{A}(t)B(t)-\dot{B}(t)A(t)=2\kappa\Gamma\xi(t) e^{-\frac{\Gamma t}{2}}\left[B(t)\int_{0}^{t}ds\xi(s)e^{\frac{\Gamma s}{2}}+2A(t)\int_{0}^{t}ds\xi(s)e^{-\frac{\Gamma s}{2}}\right]\geq 0,
\end{equation}
therefore $\gamma_{+}(t)\geq 0$ for all $t\geq 0$. In the next step one can find that
\begin{equation}
\gamma_{-}(t)=2\Gamma -4\kappa\Gamma\xi(t)e^{-\frac{\Gamma t}{2}}\frac{B(t)\int_{0}^{t}ds\xi(s)e^{\frac{\Gamma s}{2}}+2\left(A(t)-1\right)\int_{0}^{t}ds\xi(s)e^{-\frac{\Gamma s}{2}}}{B(t)}
\end{equation}
and show that
\begin{equation}\label{ineq}
B(t)\int_{0}^{t}ds\xi(s)e^{\frac{\Gamma s}{2}}+2\left(A(t)-1\right)\int_{0}^{t}ds\xi(s)e^{-\frac{\Gamma s}{2}}\leq 0,
\end{equation}
which can be written in the form
\begin{eqnarray}\label{ineq3}
\left[1-	4\kappa\Gamma  \int_{0}^{t}ds\xi(s)e^{\frac{\Gamma s}{2}}\int_{0}^{s}d\tau\xi(\tau)e^{-\frac{\Gamma \tau}{2}}+2\kappa\Gamma \int_{0}^{t}ds\xi(s)e^{\frac{\Gamma s}{2}}\int_{0}^{t}ds\xi(s)e^{-\frac{\Gamma s}{2}}\right]
\nonumber \\ \times e^{-\Gamma t}\int_{0}^{t}ds \xi(s)e^{\frac{\Gamma s}{2}}
-2\int_{0}^{t}ds\xi(s)e^{-\frac{\Gamma s}{2}}\leq 0.
\end{eqnarray}
This inequality can be proved by using
\begin{equation}\label{ineq2}
e^{-\Gamma t}\int_{0}^{t}ds\xi(s)e^{\frac{\Gamma s}{2}}\leq\int_{0}^{t}ds\xi(s)e^{-\frac{\Gamma s}{2}},
\end{equation}
which follows from
\begin{equation}\label{ineq4}
e^{-\Gamma t}\int_{0}^{t}ds\xi(s)e^{\frac{\Gamma s}{2}}\leq
e^{-\frac{\Gamma t}{2}}\int_{0}^{t}ds\xi(s)\leq
\int_{0}^{t}ds\xi(s)e^{-\frac{\Gamma s}{2}}
\end{equation}
and by applying
\begin{equation}
2\kappa\Gamma \int_{0}^{t}ds\xi(s)e^{\frac{\Gamma s}{2}}\int_{0}^{t}ds\xi(s)e^{-\frac{\Gamma s}{2}}<1+	4\kappa\Gamma  \int_{0}^{t}ds\xi(s)e^{\frac{\Gamma s}{2}}\int_{0}^{s}d\tau\xi(\tau)e^{-\frac{\Gamma \tau}{2}}.
\end{equation}
Let us notice that on both sides of the last inequality we deal with non-decreasing functions of $t$ and one can prove it by comparing values of these function at $t=0$ and their derivatives. Hence we obtain that $\gamma_{-}(t)\geq 2\Gamma$ for all $t\geq 0$.

Now we show that if (\ref{assumption}) holds, then $C(t)>0$. Let us notice that from
\begin{equation}\label{ineq6}
1-	4\kappa\Gamma \int_{0}^{t}ds\xi(s)e^{\frac{\Gamma s}{2}}\int_{0}^{s}d\tau\xi(\tau)e^{-\frac{\Gamma \tau}{2}}>0,
\end{equation}
which follows from $B(t)>0$, and
\begin{eqnarray}\label{equ1}
\int_{0}^{t}ds\xi(s)e^{\frac{\Gamma s}{2}}\int_{0}^{s}d\tau\xi(\tau)e^{-\frac{\Gamma \tau}{2}}=-\int_{0}^{t}ds\xi(s)e^{-\frac{\Gamma s}{2}}\int_{0}^{s}d\tau\xi(\tau)e^{\frac{\Gamma \tau}{2}}+\int_{0}^{t}ds\xi(s)e^{-\frac{\Gamma s}{2}}\int_{0}^{t}ds\xi(s)e^{\frac{\Gamma s}{2}}
\end{eqnarray}
we obtain
\begin{eqnarray}
1-2\kappa\Gamma \int_{0}^{t}ds\xi(s)e^{-\frac{\Gamma s}{2}}\int_{0}^{s}ds\xi(\tau)e^{\frac{\Gamma \tau}{2}}>\;\;\;\;\nonumber\\2\kappa\Gamma\left(2\int_{0}^{t}ds\xi(s)e^{-\frac{\Gamma s}{2}}\int_{0}^{t}ds\xi(s)e^{\frac{\Gamma s}{2}}-3\int_{0}^{t}ds\xi(s)e^{-\frac{\Gamma s}{2}}\int_{0}^{s}d\tau\xi(\tau)e^{\frac{\Gamma \tau}{2}}\right).
\end{eqnarray}
To check that
\begin{equation}\label{ineq1}
\int_{0}^{t}ds\xi(s)e^{-\frac{\Gamma s}{2}}\int_{0}^{t}ds\xi(s)e^{\frac{\Gamma s}{2}}\geq  \frac{3}{2}\int_{0}^{t}ds\xi(s)e^{-\frac{\Gamma s}{2}}\int_{0}^{s}d\tau\xi(\tau)e^{\frac{\Gamma \tau}{2}}
\end{equation}
we can use the fact that on both sides of the above inequality we deal with non-decreasing functions of $t$ having the same values for $t=0$ whose derivatives satisfies the inequality
\begin{equation}
\xi(t)e^{-\frac{\Gamma t}{2}}\int_{0}^{t}ds\xi(s)e^{\frac{\Gamma s}{2}}+\xi(t)e^{\frac{\Gamma t}{2}}\int_{0}^{t}ds\xi(s)e^{-\frac{\Gamma s}{2}}\geq  \frac{3}{2}\xi(t)e^{-\frac{\Gamma t}{2}}\int_{0}^{t}ds\xi(s)e^{\frac{\Gamma s}{2}}.
\end{equation}
Hence $C(t)>0$ for all $t\geq 0$. Using this results one can easily show that
\begin{equation}
\dot{C}(t)=-\frac{\Gamma}{2}C(t)-2\kappa\Gamma e^{-\Gamma t}\xi(t)\int_{0}^{t}ds \xi(s)e^{\frac{\Gamma s}{2}}<0
\end{equation}
so we see that $\gamma_{+}(t)+\gamma_{-}(t)+4\gamma_{z}(t)> 0$ is satisfied for all $t\geq 0$.

Finally let us derive
\begin{equation}
\dot{B}(t)C(t)-2B(t)\dot{C}(t)=4\kappa\Gamma \xi(t)e^{-\frac{\Gamma t}{2}} \left[B(t)e^{-\frac{\Gamma t}{2}}\int_{0}^{t}ds\xi(s)e^{\frac{\Gamma s}{2}}-C(t)\int_{0}^{t}ds\xi(s)e^{-\frac{\Gamma s}{2}}\right].
\end{equation}
To prove that this expression is less or equal to zero one can use the properties (\ref{equ1}), (\ref{ineq1}), and (\ref{ineq4}). This property leads to $\gamma_{z}(t)\leq 0$ for all
$t\geq 0$ and ends the proof. \hfill $\Box$

\section{ Proof to Proposition \ref{PRO-5}}\label{max_exp}

Assuming $\rho(0) = |g\rangle \langle g|$ one has

\begin{equation}\label{}
P_e(\alpha,\Delta_{0},t) = \frac{4 \kappa\alpha \Gamma^2}{(1-\alpha)^2\Gamma^2+4\Delta_{0}^2} \left(e^{-\Gamma t}-2e^{-\frac{(1+\alpha)\Gamma t}{2}}\cos  \Delta_{0} t+e^{-\alpha\Gamma t}\right).
\end{equation}
Introducing $\chi$ such that $\Gamma t/2 = \chi$ one finds that the function

\begin{equation}\label{}
P_e= f(\alpha,\Delta_{0}, \chi) =  \frac{4 \kappa\alpha \Gamma^2}{(1-\alpha)^2\Gamma^2+4\Delta_{0}^2} \left(e^{-2\chi}-2e^{-(1+\alpha)\chi}\cos  \Delta_{0} t+e^{-\alpha\chi}\right).
\end{equation}
attains its local maximum for $\alpha=1$, $\Delta_{0}=0$, and $\chi=1$ which corresponds to $t = 2/\Gamma$. Finally,

\begin{equation}\label{}
\lim_{\alpha \to 1} f(\alpha,0,1) = \frac{4\kappa}{e^2} ,
\end{equation}
which ends the proof.  \hfill $\Box$

\section{formulae for $\gamma_{+}$, $\gamma_{-}$, and $\gamma_{z}$ for $\Delta_{0}=0$}\label{Appengammas}

If $\alpha\neq 1$,  then for (\ref{A_resonant})-(\ref{C_resonant}) we obtain
\begin{align}
\gamma_+(t)&=16 \alpha  \Gamma  \kappa  e^{-\frac{1}{4} (\alpha +1) \Gamma  t} \sinh \left[\frac{1}{4} (\alpha -1) \Gamma  t\right]\frac{ \left(\alpha ^2-16 \kappa -1\right) \sinh \left[\frac{\alpha  \Gamma  t}{2}\right]+\left(\alpha ^2-1\right) \cosh \left[\frac{\alpha  \Gamma  t}{2}\right]+16 \alpha  \kappa  \sinh \left[\frac{\Gamma  t}{2}\right]}{(\alpha -1) \left\{8\kappa (1-\alpha)  +16 \alpha  \kappa  e^{\frac{1}{2} (\alpha +1) \Gamma  t}+(\alpha +1) (\alpha -8 \kappa -1) e^{\alpha  \Gamma  t}\right\}}, \\
\gamma_-(t)&=-\gamma_+(t)+ \frac{2 (\alpha +1) \Gamma  \left\{8 \kappa (1-\alpha)  +8 \alpha  \kappa  e^{\frac{1}{2} (\alpha +1) \Gamma  t}+(\alpha -8 \kappa -1) e^{\alpha  \Gamma  t}\right\}}{8 \kappa (1-\alpha)  +16 \alpha  \kappa  e^{\frac{1}{2} (\alpha +1) \Gamma  t}+(\alpha +1) (\alpha -8 \kappa -1) e^{\alpha  \Gamma  t}}, \\
\gamma_z(t)&=-\frac{1}{4}(\gamma_+(t)+\gamma_-(t))-\frac{\Gamma  \left\{-8 \alpha  (\alpha +2) \kappa  e^{\frac{\alpha  \Gamma  t}{2}}+4 (\alpha +1) (2 \alpha +1) \kappa  e^{\frac{\Gamma  t}{2}}-(\alpha -1) (\alpha -4 \kappa +1) e^{\left(\alpha +\frac{1}{2}\right) \Gamma  t}\right\}}{16 \alpha  \kappa  e^{\frac{\alpha  \Gamma  t}{2}}-8 (\alpha +1) \kappa  e^{\frac{\Gamma  t}{2}}+2 (\alpha -1) (\alpha -4 \kappa +1) e^{\left(\alpha +\frac{1}{2}\right) \Gamma  t}}.
\end{align}

For $\alpha=1$, from (\ref{A_resonant_1})-(\ref{C_resonant_1}) we get
\begin{align}
\gamma_+(t)&=\frac{4 \Gamma ^2 \kappa  t \Big\{2 \kappa  (\Gamma  t+2)+e^{\Gamma  t} \big[2 \kappa  (\Gamma  t-2)-1\big]\Big\}}{4 \kappa  e^{\Gamma  t}+e^{2 \Gamma  t} (4 \kappa  (\Gamma  t-1)-1)} \\
\gamma_-(t)&=-\gamma_+(t)+\frac{2 \Gamma  \Big\{8 \kappa +e^{\Gamma  t} \big[4 \kappa  (\Gamma  t-2)-1\big]\Big\}}{4 \kappa +e^{\Gamma  t} (4 \kappa  (\Gamma  t-1)-1)} \\
\gamma_z(t)&=-\frac{1}{4}(\gamma_+(t)+\gamma_-(t))+\frac{\Gamma}{2}   \left[\frac{4 \Gamma  \kappa  t}{2 (\kappa +\Gamma  \kappa  t)+(1-2 \kappa ) e^{\Gamma  t}}+1\right]
\end{align}
Hence one can find the limits (\ref{Ia}) and (\ref{IIa}).

\section{ Proof to Proposition \ref{PRO-5}}\label{AppenD}

One can rewrite (\ref{B_resonant}) as
\begin{equation}
B(t)
= \frac{16\kappa\alpha}{\alpha^2-1} e^{-\frac{1+\alpha}2 \Gamma t} + \left( 1 - \frac{8\kappa}{\alpha-1} \right) e^{-\Gamma t} - \frac{8\kappa}{\alpha+1} e^{-(1+\alpha)\Gamma t} \label{B_resonant_proof}
\end{equation}
If $\alpha < 1$, then we rewrite the above as:
\begin{equation}
B(t) =
e^{-\frac{1+\alpha}2 \Gamma t} \left(
\frac{16\kappa\alpha}{\alpha^2-1} + \left( 1 - \frac{8\kappa}{\alpha-1} \right) e^{-\frac{1-\alpha}2 \Gamma t} - \frac{8\kappa}{\alpha+1} e^{-\frac{1+\alpha}2 \Gamma t}
\right)
\end{equation}
The limit of the second factor is $16\kappa\alpha/(\alpha^2-1) < 0$, hence $B(t)$ is negative for $t$ large enough. Because $B(0) = 1$, there exists $t$ such that $B(t)=0$.

If $\alpha \in (1, 8\kappa+1)$, then rewriting (\ref{B_resonant_proof}) as:
\begin{equation}
B(t) =
e^{-\Gamma t}
\left(
\frac{16\kappa\alpha}{\alpha^2-1} e^{-\frac{\alpha-1}2 \Gamma t} + \left( 1 - \frac{8\kappa}{\alpha-1} \right) - \frac{8\kappa}{\alpha+1} e^{-\alpha\Gamma t}
\right)
\end{equation}
one can notice that the limit of the second factor is $1 - 8\kappa/(\alpha-1)$ which now is negative due to assumption about the range of $\alpha$ and again there exists $t$ such that $B(t)=0$.

If $\alpha = 1$, then the second factor in (\ref{B_resonant_1}) is equal 1 for $t=0$ and it is unbounded from below, hence there exists $t$ such that $B(t)=0$ . Hence for the whole range $\alpha < 8\kappa+1$ the dynamics is not invertible.

If $\alpha \geq 8\kappa+1$, then $1 - 8\kappa/(\alpha-1) \geq 0$ and estimating (\ref{B_resonant_proof}) as follows:
\begin{align}
B(t)
& =
\frac{16\kappa\alpha}{\alpha^2-1} e^{-\frac{1+\alpha}2 \Gamma t} + \left( 1 - \frac{8\kappa}{\alpha-1} \right) e^{-\Gamma t} - \frac{8\kappa}{\alpha+1} e^{-(1+\alpha)\Gamma t}
\nonumber \\
& \ge
\frac{16\kappa\alpha}{\alpha^2-1} e^{-\frac{1+\alpha}2 \Gamma t} + \left( 1 - \frac{8\kappa}{\alpha-1} \right) e^{-\frac{1+\alpha}2 \Gamma t} - \frac{8\kappa}{\alpha+1} e^{-(1+\alpha)\Gamma t}
=
\left( 1 + \frac{8\kappa}{\alpha+1} \right) x - \frac{8\kappa}{\alpha+1} x^2
\end{align}
where $x = \exp(-(1+\alpha)\Gamma t/2) \in (0,1]$, one can observe that the above expression is positive for $x \in (0,1]$, hence $B(t)$ is positive for all $t$.

Finally, using the result from Appendix \ref{AppenC} i.e  whenever $B(t)>0$ then $C(t)>0$, we see that for the range $\alpha \geq 8\kappa + 1$, we get $C(t)>0$, and we conclude that the dynamics is invertible.

\end{document}